\documentclass[twocolumn,showpacs,preprintnumbers,
amsmath,amssymb,aps,prd,nofootinbib,superscriptaddress,
eqsecnum]{revtex4}
\usepackage{graphicx,color}
\makeatletter
\renewcommand{\p@subsection}{}
\makeatother

\newcommand{\Slash}[1]{\ooalign{\hfil/\hfil\crcr$#1$}}
\begin{document}
\title{
Chiral phase transition in the presence of spinodal decomposition
}
\author{C. Sasaki}
\affiliation{%
Technische Universit\"at M\"unchen,  D-85748 Garching, Germany }
\author{B. Friman}
\affiliation{%
Gesellschaft f\"ur Schwerionenforschung, GSI,  D-64291 Darmstadt,
Germany}
\author{K. Redlich}
\affiliation{%
Institute of Theoretical Physics, University of Wroclaw, PL--50204 Wroc\l aw, Poland}
\affiliation{%
Institute f\"ur Kernphysik, Technische Universit\"at Darmstadt, D-64289 Darmstadt,
Germany}

\date{\today}
\begin{abstract}
The thermodynamics  of a first order  chiral phase transition is considered in the
presence of spinodal phase separation within the Nambu-Jona-Lasinio (NJL) model. The
properties of the basic thermodynamic observables in the  coexistence phase are discussed
for zero and non-zero quark masses. We focus on observables that probe the chiral phase
transition. In particular, the behavior of the specific heat and entropy as well as
charge fluctuations are calculated and analyzed. We show that the specific heat and
charge susceptibilities diverge at the isothermal spinodal lines. We  determine the
scaling behavior and compute the critical exponent $\gamma$ of the net quark number
susceptibility at the isothermal spinodal lines within the  NJL model and the
Ginsburg-Landau theory. We show that in the chiral limit the critical exponent
$\gamma=1/2$ at the tricritical point as well as along the isothermal spinodal lines. On
the other hand, for finite quark masses the critical exponent at the spinodal lines,
$\gamma=1/2$, differs from that at the critical end point, $\gamma=2/3$, indicating a
change in the universality class. These results are independent of the particular choice
of the chiral Lagrangian and should be common for all mean field approaches.

\end{abstract}
\setcounter{footnote}{0}
\maketitle

\section{Introduction}
\label{sec:int}

A first order phase transition is intimately linked with the existence of a convex
anomaly in thermodynamic pressure \cite{ran}, which can be probed only in non-equilibrium
states. There is an interval of energy and/or baryon number density where the derivative
of pressure with respect to the volume of the system is positive. This anomalous behavior
defines a region of instability in the ($T,n_q)$-plane with $n_q$ being the net quark
number density. The region is bounded by the spinodal lines, where the pressure
derivative with respect to volume vanishes. The derivative taken at constant temperature
or entropy defines the isothermal and isentropic spinodal lines, respectively. Spinodal
decomposition is thought to play a dominant role in the dynamics of low energy nuclear
collisions in the regime of the first order nuclear liquid-gas
transition~\cite{ran,heiselberg}. Furthermore, the consequences of spinodal decomposition
in the connection with the chiral and deconfinement phase transitions in heavy ion
collisions have been discussed in Refs.~\cite{ran,gavin,rans,ranhi,heiselberg,polony}. It
has been argued that in the region of phase coexistence, a spinodal phase separation
leads to a significant enhancement of baryon \cite{gavin} and strangeness fluctuations
\cite{rans}. Studies of non-equilibrium dynamics in the linear sigma model have shown
that significant density inhomogeneities may be dynamically induced in a first-order
chiral phase transition~\cite{paech}. In a recent paper it was shown that the spinodal
instability is connected with divergent fluctuations of the net quark number, $\langle
N_q N_q\rangle - \langle N_q\rangle \langle N_q\rangle=V T \chi_q$, along the isothermal
spinodal lines~\cite{SFR:spinodal}. Consequently, the quark number susceptibility
$\chi_q$  diverges not only at the critical end point (CEP) in the  QCD phase diagram but
also at the (non-equilibrium) first order phase transition. The close relation of the
singularities at the CEP with those at the spinodal lines is a consequence of the
thermodynamic relations connecting the compressibility of the system with the quark
number susceptibility. In this paper we explore the critical behavior and properties of
the fluctuations of conserved charges and the specific heat in the presence of spinodal
phase decomposition using the  Nambu-Jona-Lasinio (NJL) model. In particular, we study
the influence of an explicit chiral symmetry breaking  through the finite value of the
quark mass on the thermodynamics across the off-equilibrium first order transition. The
critical exponents of $\chi_q$ at the CEP and along the spinodal lines are computed in
the NJL model and in the Ginzburg-Landau effective theory. In the chiral limit, the
critical exponents at the CEP and at the spinodal lines are identical, while for finite
vales of the quark mass, the exponents differ. Our calculations are done within the
mean-field approximation to the NJL model. However, the results are generic and should
apply to all QCD-like chiral models. We stress at this point that for non-universal
quantities, like the position of the CEP, one can expect only qualitative agreement
between different models. On the other hand, for universal quantities, like the critical
exponents, all models in the same universality class yield the same result~\footnote{The
critical exponents obtained in the mean-field approximation are in general modified by
fluctuations.}.

The paper is organized as follows: In Section~\ref{sec:NJL} we explore the general
features of the spinodal instabilities within the NJL model. In Section~\ref{sec:sus} we
discuss  the critical behavior of the quark number susceptibility in the mean field
approximation. The critical exponents of $\chi_q$ are computed in
Ginzburg-Landau theory in Section~\ref{sec:GL}. Concluding remarks are presented in
Section~\ref{sec:sum}.

\begin{figure*}
\begin{center}
\includegraphics[width=8.7cm]{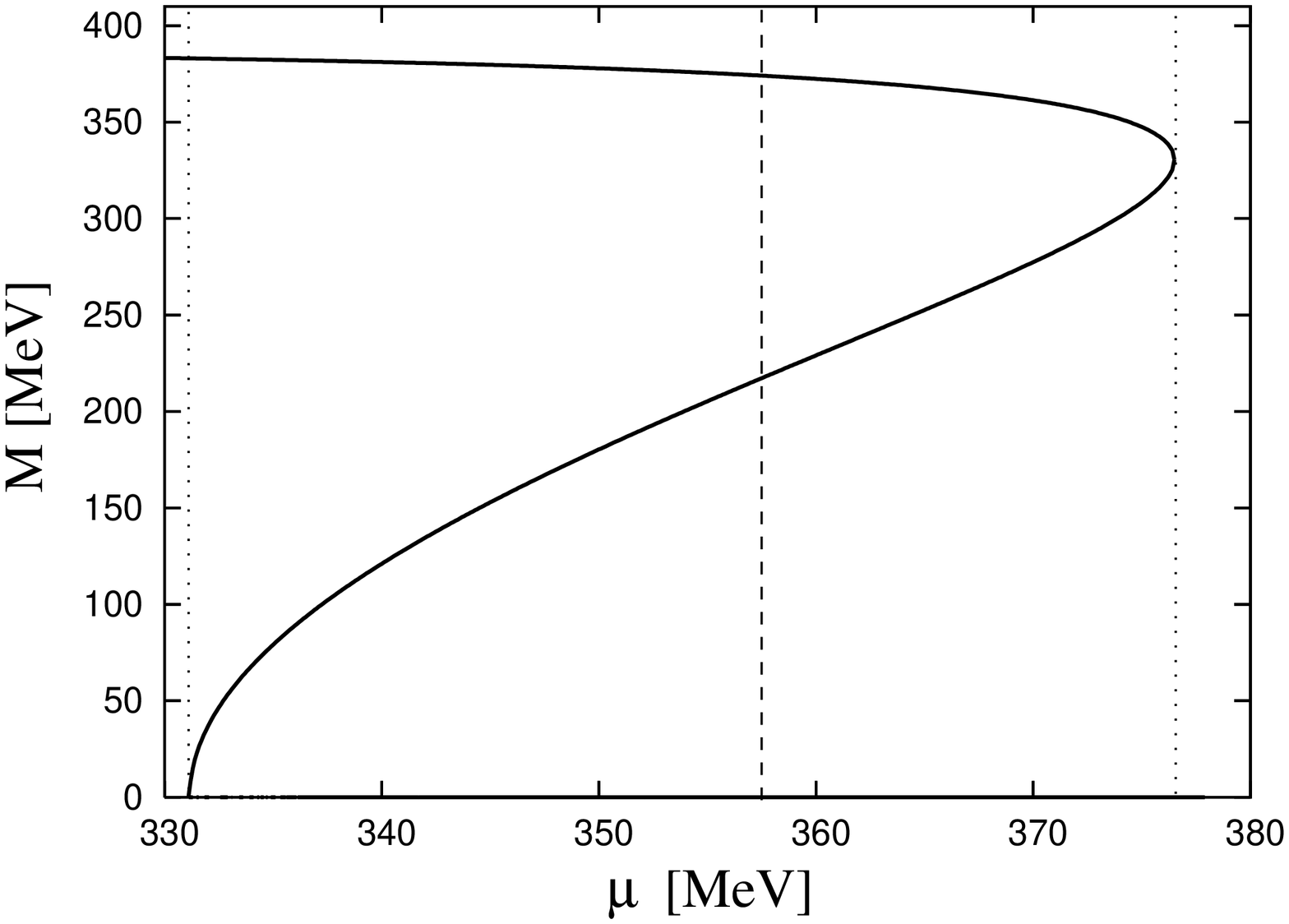}
\includegraphics[width=8.7cm]{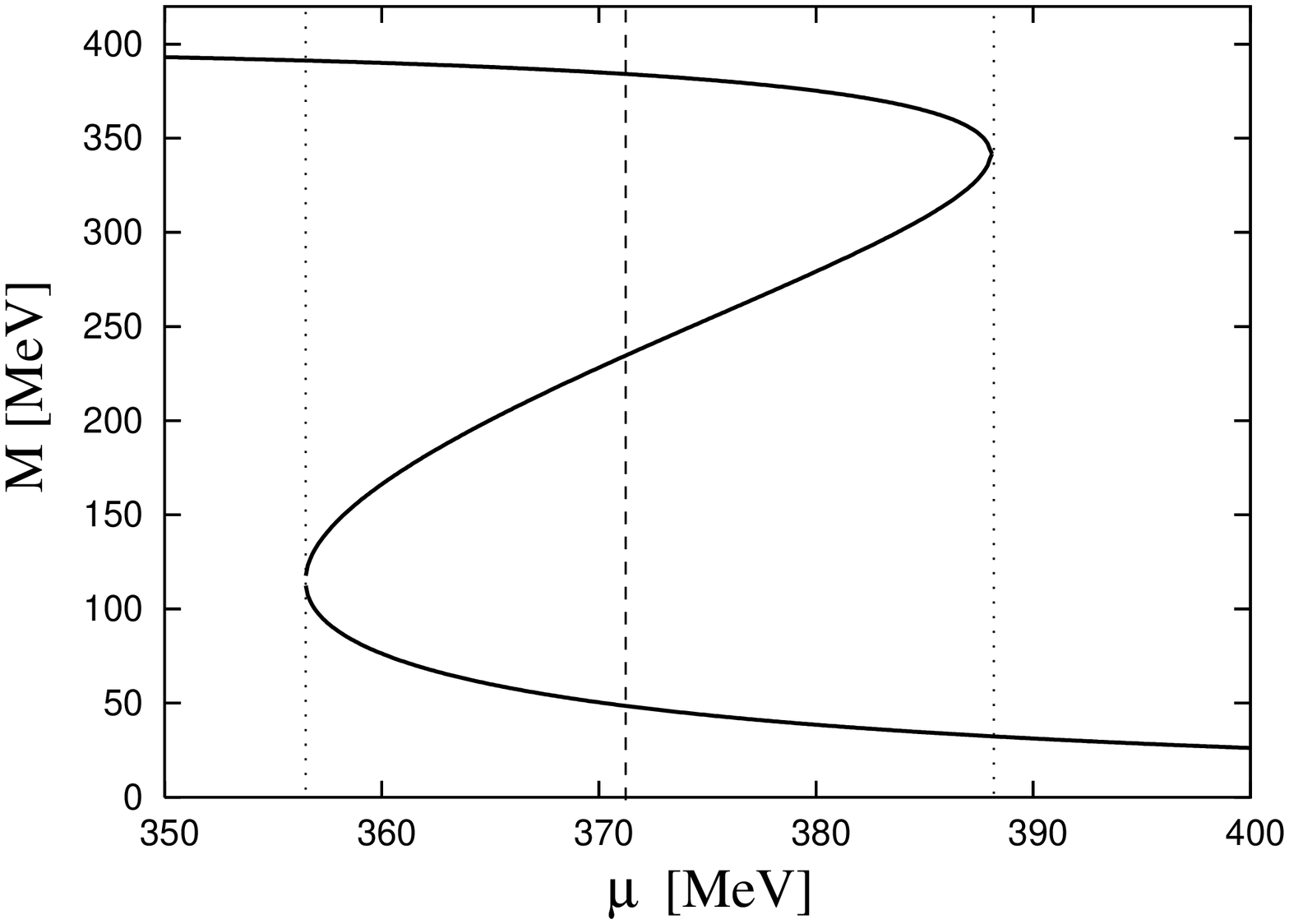}
\caption{ The dynamical quark mass at fixed temperature ($T=30$ MeV) as a function of the
quark  chemical potential in the chiral limit (left panel) and for a finite quark mass
$m=5.6$ MeV (right panel). The dashed line indicates the equilibrium first order phase
transition while the dotted lines show the isothermal spinodal points.
 }
\label{mass}
\end{center}
\end{figure*}
\begin{figure*}
\begin{center}
\includegraphics[width=8.7cm]{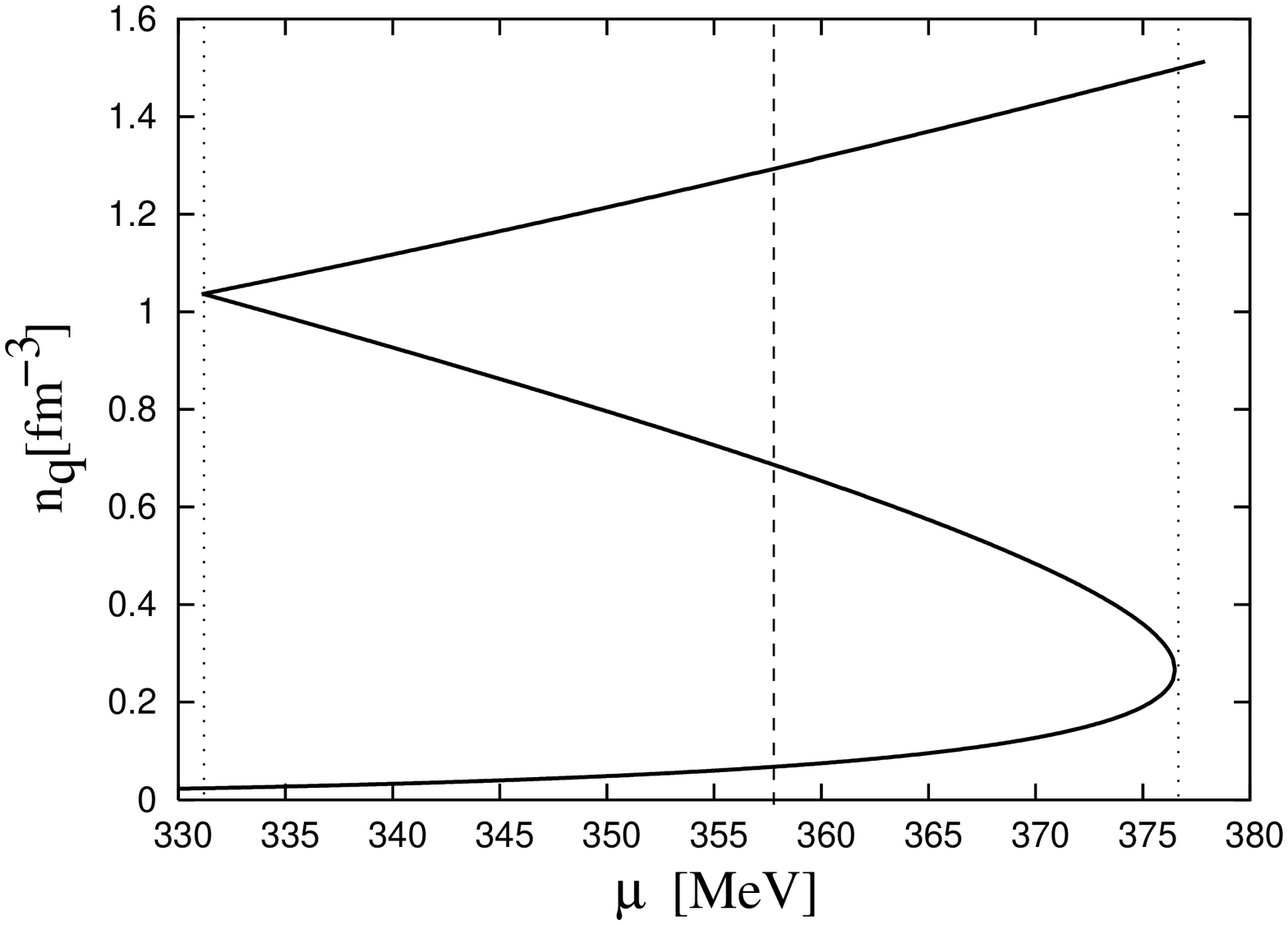}
\includegraphics[width=8.7cm]{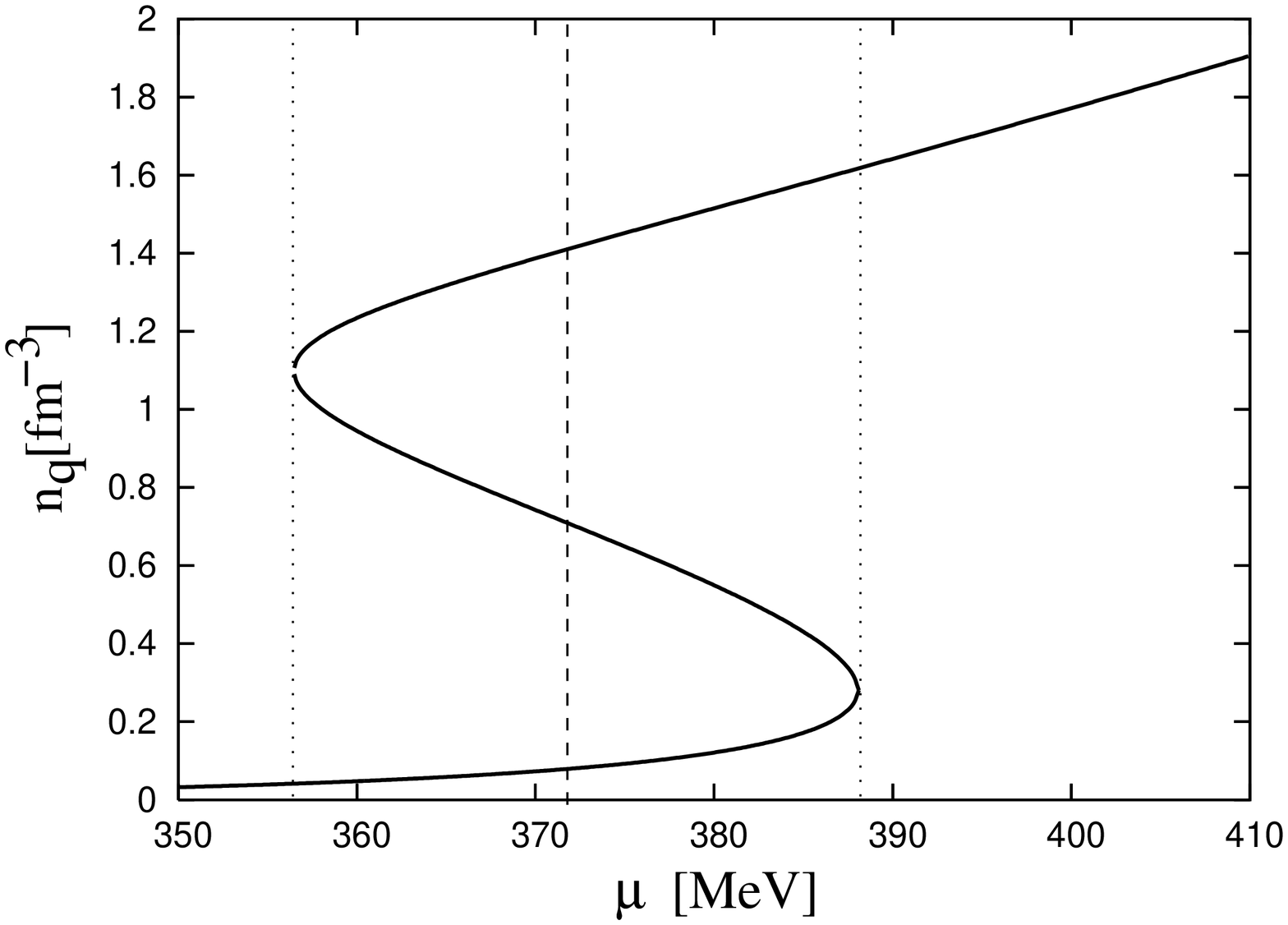}
\caption{ The net  quark number density at $T=30$ MeV computed in the chiral limit (left
panel) and for a finite quark mass $m=5.6$ MeV (right panel). The dashed and dotted lines
have the same meaning as in Fig.~\ref{mass}.
 }
\label{nq}
\end{center}
\end{figure*}
 \setcounter{equation}{0}
\section{Effective Chiral  Model and Spinodal instabilities}
\label{sec:NJL}

We study the thermodynamics of the chiral phase transition in the presence of spinodal
 instabilities within the Nambu--Jona-Lasinio (NJL) model~\cite{nambu}. The model describes
 the effective interactions of quarks preserving the chiral symmetry of the massless QCD Lagrangian.
 We include quark interactions only in the scalar and isoscalar sectors. In this case the NJL Lagrangian
  for two quark flavors, three colors and for finite quark chemical potentials
   has the form~\cite{kunihiro,hatsuda,sexy,review}:
\begin{align}
\label{eq1}
{\mathcal L} = \bar{\psi}( i\Slash{\partial} -m )\psi {}+ \bar{\psi}\mu_q\gamma_0\psi {}+ G_S
 \Bigl[ \bigl( \bar{\psi}\psi \bigl)^2 + \bigl( \bar{\psi}i\vec{\tau}\gamma_5\psi \bigl)^2  \Bigr]\,,
 \end{align}
where $m = \mbox{diag}(m_u, m_d)$ is  the current quark mass, $\mu_q = \mbox{diag} (\mu_u, \mu_d)$ is
 the  quark chemical potential  and $\vec{\tau}$ are Pauli matrices. The strength of the interactions
 among the constituent quarks  is controlled by the coupling $G_S \Lambda^2 = 2.44$ with the three
  momentum cut-off $\Lambda=587.9$ MeV, introduced to regulate the ultraviolet divergences
   \cite{review}. The parameters of the model are fixed so as to reproduce  the pion mass
    and decay constant in vacuum for $m_u=m_d=5.6$ MeV.

In the mean field approximation the thermodynamics of the NJL model, for an isospin
symmetric system, is obtained from the thermodynamic potential~\cite{review}:
\begin{align}\label{eq2}
& \Omega (T,\mu;M)/V = \frac{(M-m)^2}{4G_S} {}- 12 \int\frac{d^3p}{(2\pi)^3}
\Bigl[E(\vec{p}\,)
\nonumber\\
&
{}- T\ln ( 1-n^{(+)}(\vec{p},T,\mu)
)\Bigr.
{}-\Bigl.  T\ln (1-n^{(-)}(\vec{p},T,\mu) \Bigr]\,,
\end{align}
where  $M = m- 2G_S\langle \bar{\psi}\psi \rangle$ is the dynamical quark mass,
 $E(\vec{p}\,) = \sqrt{\vec{p}^{\,2} + M^2}$ is its energy and $n^{(\pm)}(\vec{p},T,\mu) =
  \Bigl( 1 + \exp\bigl[ (E(\vec{p}\,) \mp \mu)/T \bigr] \Bigr)^{-1}$ is the particle/antiparticle
  distribution function. The quark chemical potential $\mu$ is expressed as an
    average $\mu=\mu_u = \mu_d$. The dynamical quark mass $M$ in Eq. (\ref{eq2})
    is obtained self-consistently from the stationarity
    condition  ${\partial\Omega}/{\partial M} = 0$, which implies:
\begin{align}\label{eq3}
M = m+24 G_S \int\frac{d^3 p}{(2\pi)^3} \frac{M}{E} \Bigl[ 1 - n^{(+)} - n^{(-)}
\Bigr]\,.
\end{align}

\begin{figure*}
\begin{center}
\includegraphics[width=8.7cm]{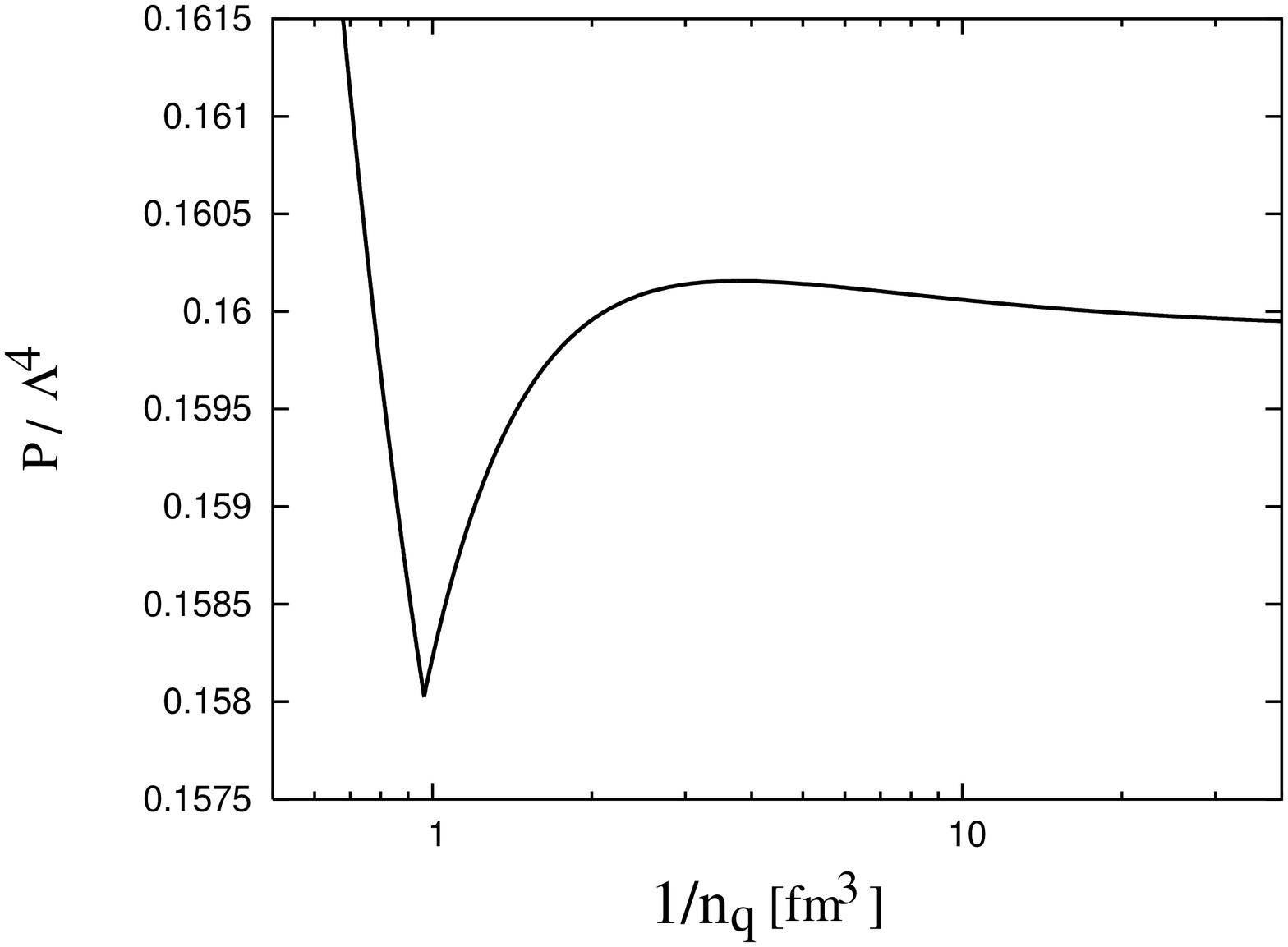}
\includegraphics[width=8.7cm]{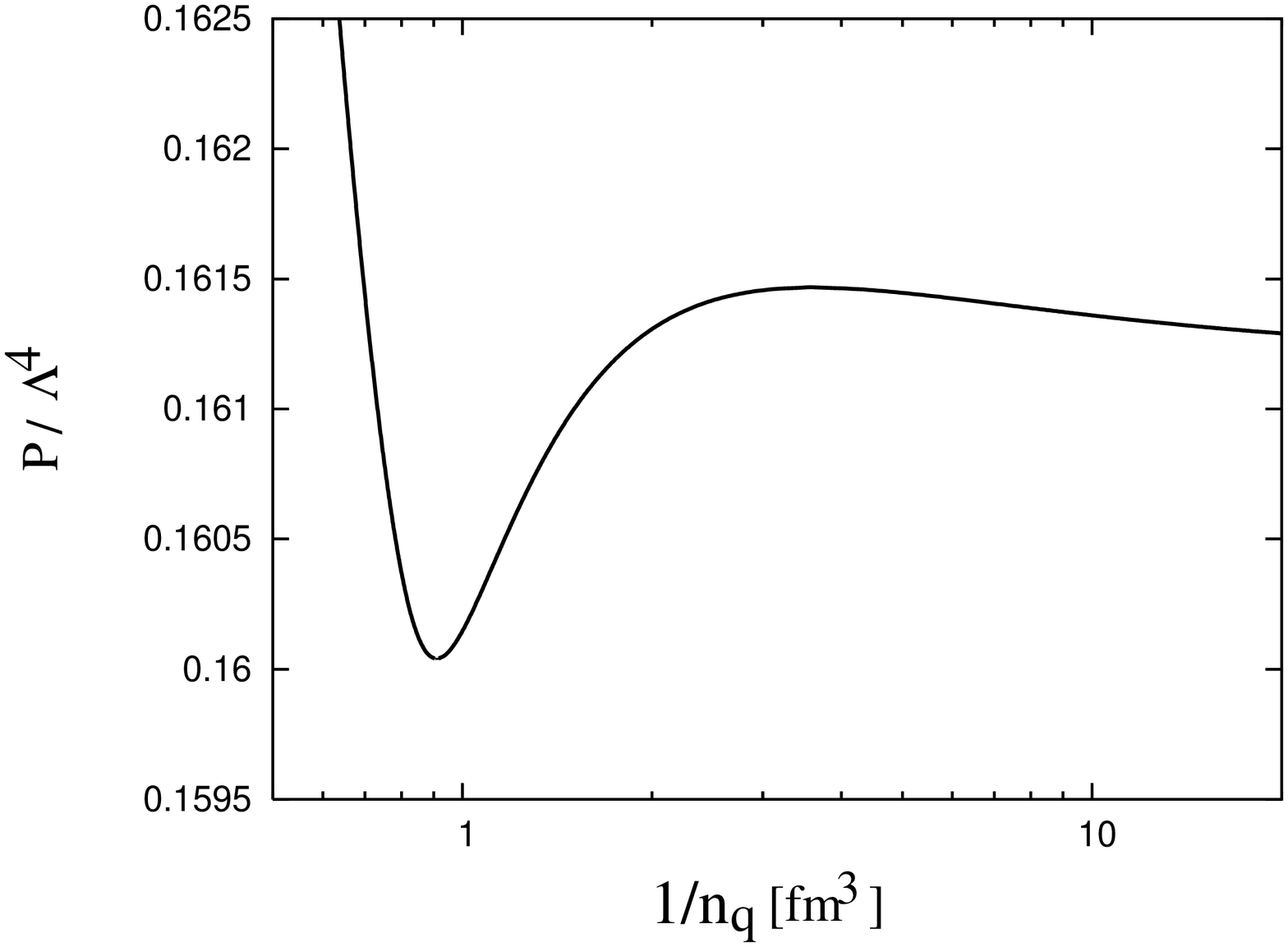}
\caption{ In the left panel we show the pressure as a function of inverse quark number density for fixed
temperature, $T=30$ MeV obtained in the chiral limit, while in the
right panel we show the results for a finite quark mass, $m=5.6$ MeV.
 }
\label{pv}
\end{center}
\end{figure*}

\begin{figure*}
\begin{center}
\includegraphics[width=8.7cm]{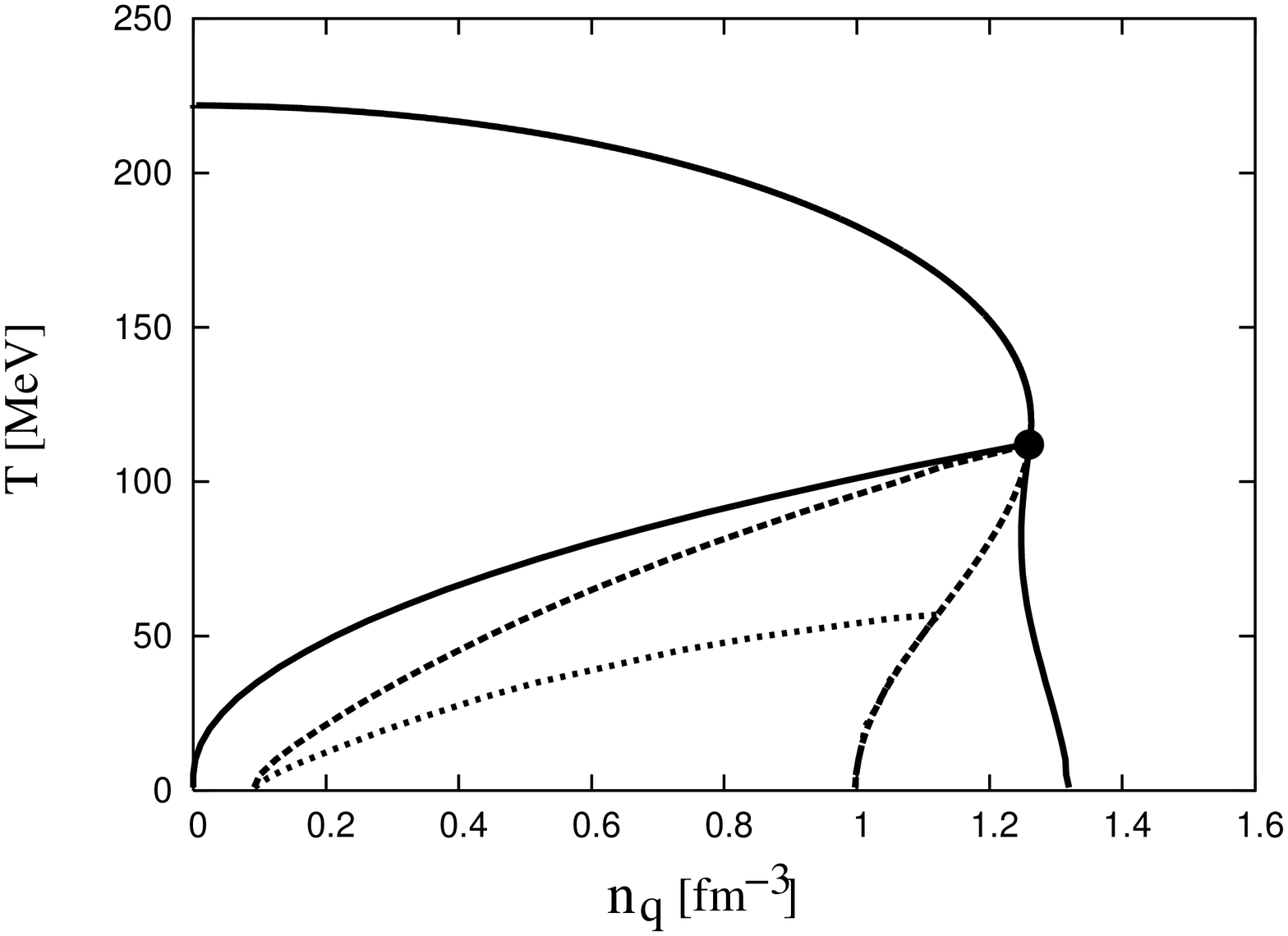}
\includegraphics[width=8.7cm]{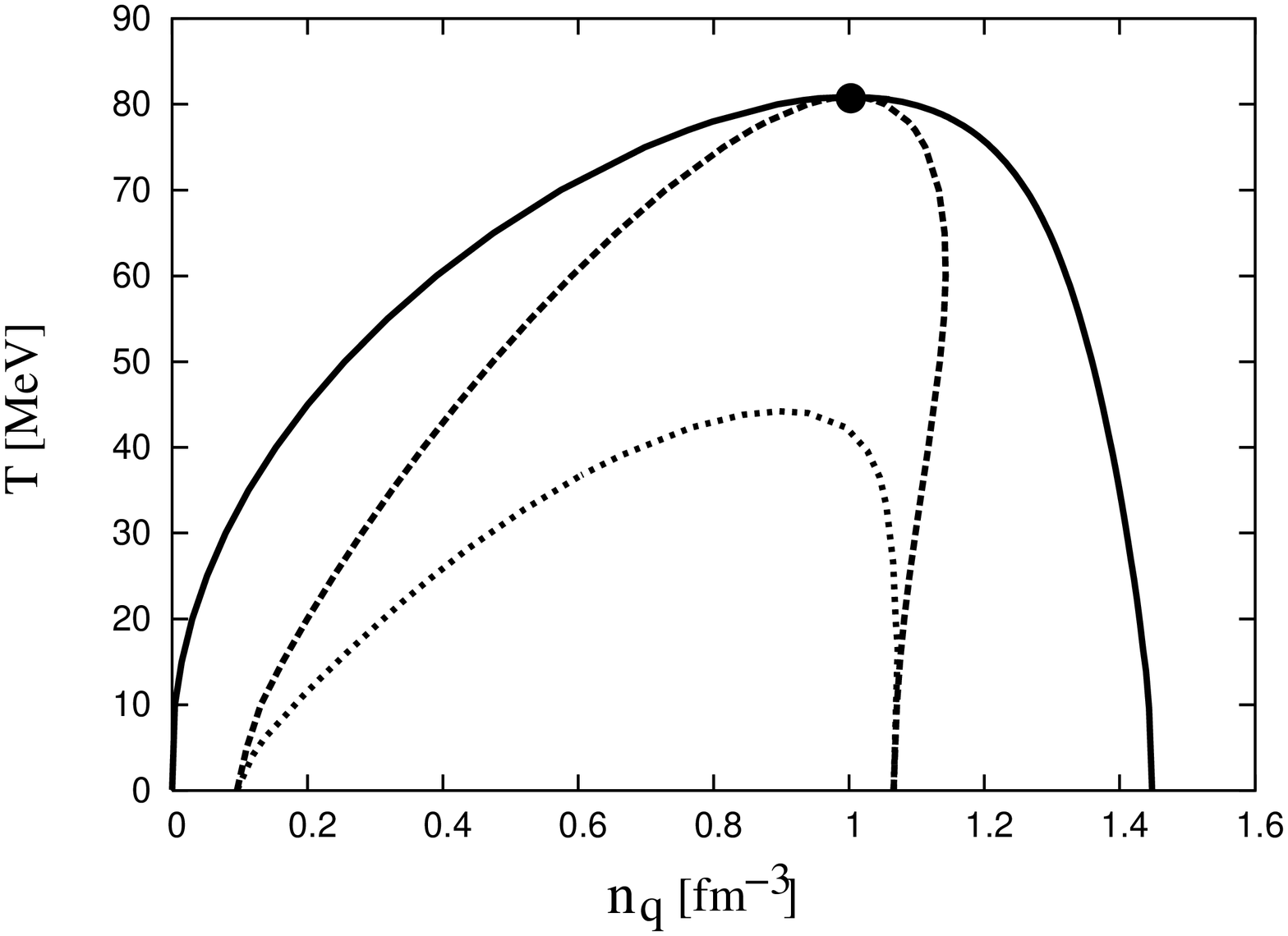}
\caption{ The phase diagram of the NJL model in $(T,n_q)$-plane.
The left figure corresponds to the chiral limit. The dot indicates
the position of the TCP and the full-line above the TCP shows the
second order transition line. The full-lines below the TCP represent the
phase boundaries of the first order chiral transition. Finally, the
 dashed (dotted) lines show the isothermal (isentropic) spinodal lines.
  The right figure was computed with a
finite   quark mass $m=5.6$ MeV. Above the CEP, indicated by a dot, there is cross-over transition. }
\label{phase}
\end{center}
\end{figure*}
The thermodynamic pressure of the system is given by $P=-\Omega/V$,
 and the net quark number density $n_q$ is given by
\begin{align}\label{eq4}
n_q &= \frac{\partial P}{\partial \mu}= 12 \int\frac{d^3 p}{(2\pi)^3} \Bigl[ n^{(+)} - n^{(-)} \Bigr]\,.
\end{align}
Thus, in the mean-field approximation the net quark density $n_q$ has
the same  structure as in a non-interacting gas of massive fermions, albeit with a
 $(T,\mu)$-dependent effective mass $M(T,\mu)$.
The dynamical quark mass $M$ is an order parameter of the chiral phase transition.
In the left panel of Fig.~\ref{mass} we show $M$ computed at fixed temperature $T=30$ MeV as
a function of the quark chemical potential, in the chiral limit, i.e., for vanishing current
quark masses. The behavior of $M$ is typical for an order parameter in a first order
phase transition. There is a range of values of the chemical potential where there are several
solutions to the gap equation. There are stable/metastable solutions, corresponding to
absolute/local minima of the thermodynamic potential and an unstable solution, corresponding to
a maximum. The equilibrium transition from the chirally broken to the symmetric phase
is obtained when the two minima are degenerate. This corresponds to the so called Maxwell
construction. The location of the equilibrium transition is indicated in Fig.~\ref{mass} by
the dashed line. A non-zero current quark mass in the NJL Lagrangian explicitly breaks the chiral
symmetry. Consequently, in this case, the dynamical
mass $M$ is not a true order parameter and it never vanishes. The dependence of $M$ on
the quark chemical potential is shown in the right panel of Fig.~\ref{mass}. The
characteristic dependence of $M$ on the chemical potential at a first order transition
is, as will be discussed below, reflected in various thermodynamic quantities.

\begin{figure*}
\begin{center}
\includegraphics[width=8.7cm]{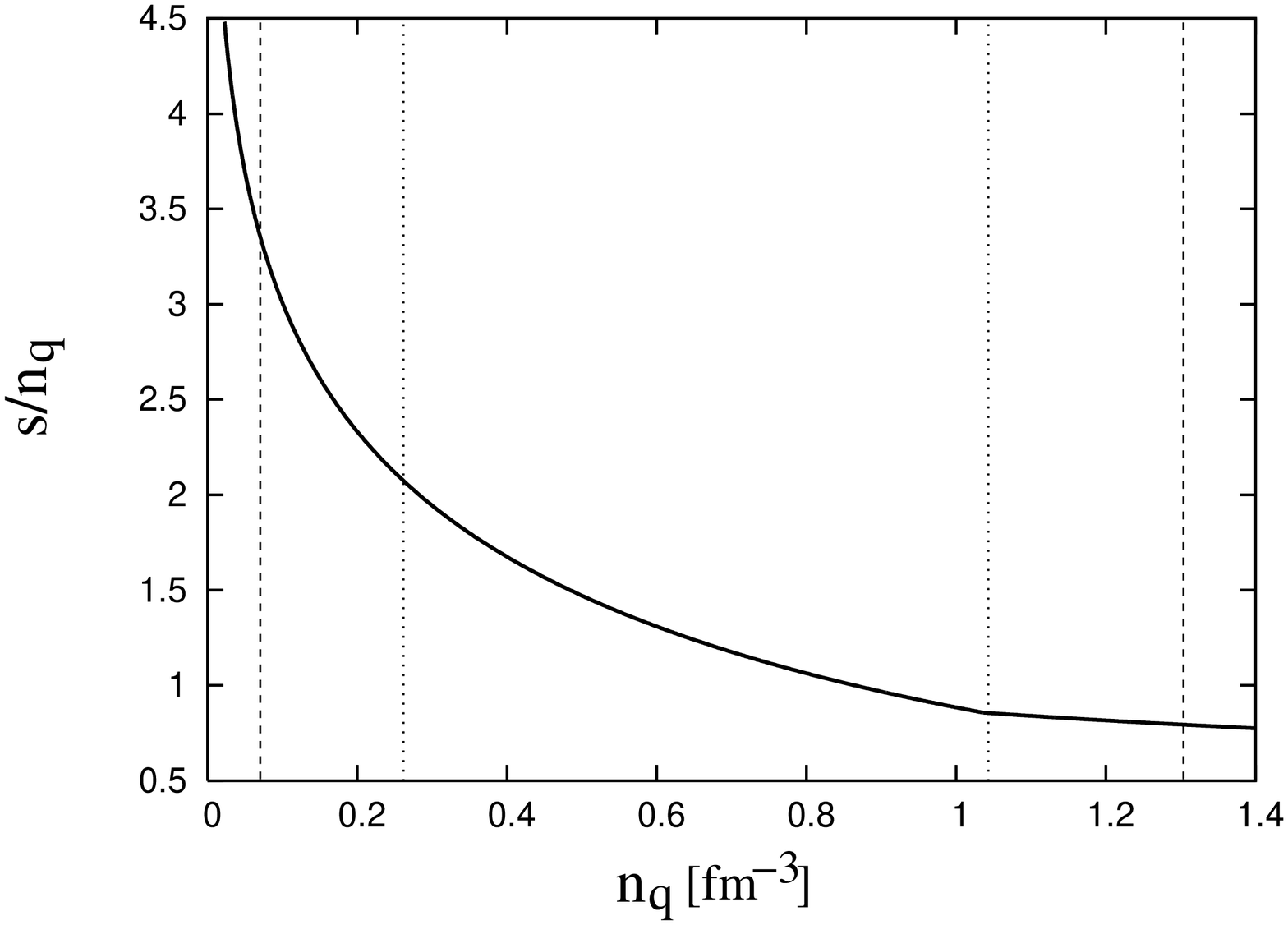}
\includegraphics[width=8.7cm]{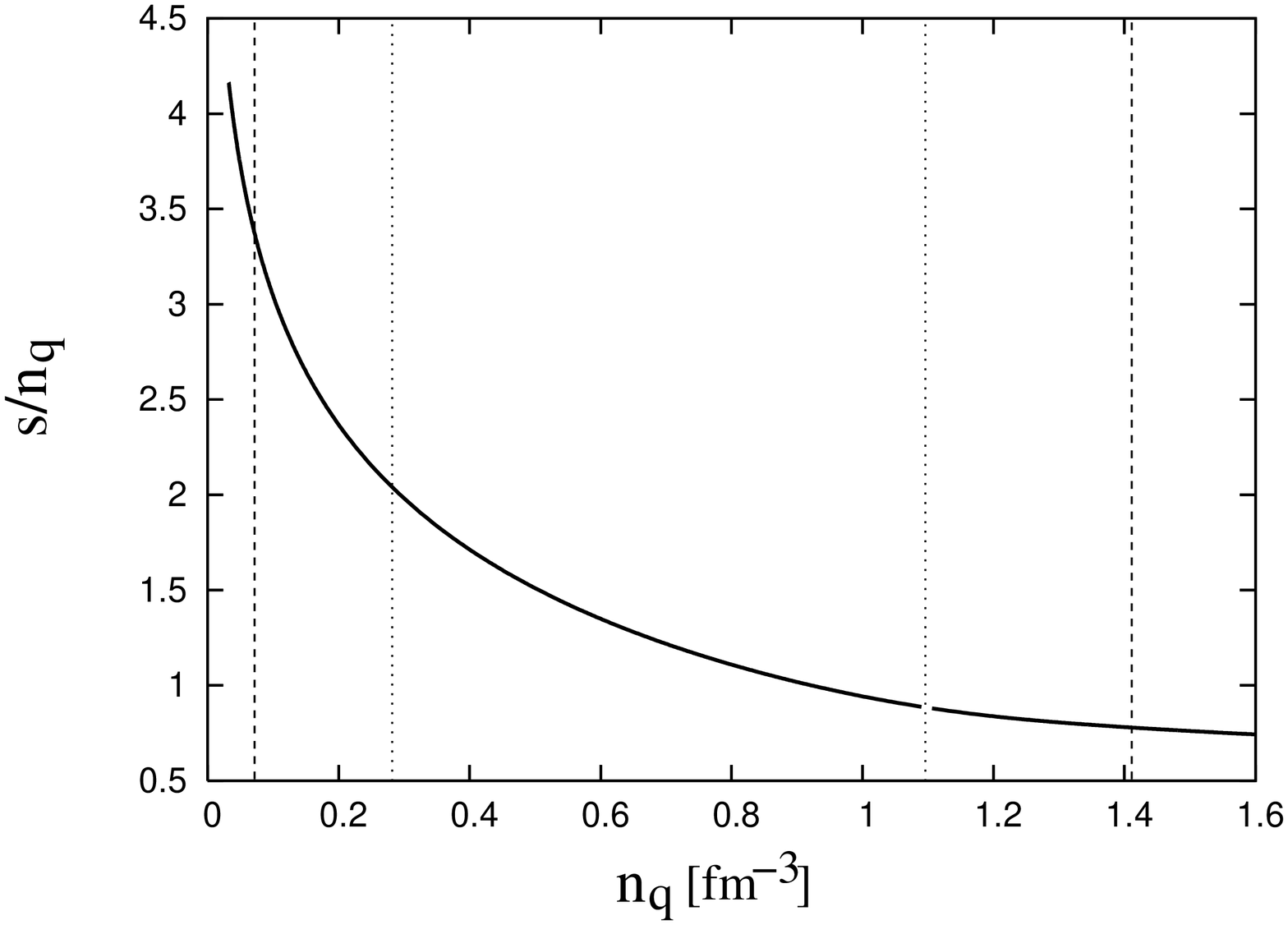}
\caption{ The ratio of the entropy to the net quark number density, i.e.,
 the entropy per quark, at fixed temperature as a function of the net quark
 density in the chiral limit (left panel) and for a finite quark mass $m=5.6$
 MeV (right-panel). The dashed and dotted lines have the same meaning as in Fig.~\ref{mass}.
 }
\label{sb}
\end{center}
\end{figure*}

In Fig.~\ref{nq} we show the net quark number density  at fixed temperature as a function
of $\mu$ for vanishing as well as for a finite current quark mass. As expected, the
non-monotonic dependence of the mass $M$ on the chemical potential results in
corresponding structure in $n_q$. However, there is an notable difference in the
properties of $n_q$ in the chiral limit and for a finite quark masses. In the latter
case, $n_q$ is differentiable at all values of $\mu$, while in the chiral limit $n_q$
exhibits a cusp at the point where the dynamical quark mass vanishes. This has important
consequences for observables obtained from $n_q$ through differentiation with respect to
thermal parameters, like e.g. the  quark number or mixed susceptibilities. The
thermodynamic pressure as a function of the volume per net quark, $(1/n_q)$, is shown in
Fig.~\ref{pv} at fixed temperature. Independently of the current quark mass, the
qualitative structure of the pressure is that of a van-der-Waals' equation of state. The
unstable solution of the gap equation corresponds to the part of the equation of state
where the volume derivative of the pressure is positive, i.e., to the mechanically
unstable region. This region is bounded by the spinodal points located at the minimum and
maximum of the pressure, respectively. Outside of this region the system is mechanically
stable~\footnote{More precisely, the system is absolutely stable outside the coexistence
region, while within this region but outside the spinodal lines it is metastable, i.e.,
stable only against small amplitude fluctuations.}. The volume derivative of the pressure
depends on which quantity is kept constant, the temperature or the entropy. If the volume
derivative of $P$ exists, then the isothermal and isentropic spinodal lines are defined
by zeroes of the corresponding derivatives
\begin{equation}\label{eq5}
\left( \frac{\partial P}{\partial V}\right)_T=0
\qquad {\rm or} \qquad
\left( \frac{\partial P}{\partial V}\right)_S=0\,.
\end{equation}
From Fig. {\ref{pv}}, it is clear that for a finite quark mass $m$, the derivative of the
thermodynamic pressure is well defined everywhere. However, in the chiral limit the
pressure exhibits a cusp rather than a regular minimum, at the point where the pressure
derivative changes sign on the high density side. Consequently, in this case equation
(\ref{eq5}) can not be used to determine the position of the spinodal line. Instead, the
corresponding branch of the isothermal or isentropic spinodal line is defined by a change
in sign of the pressure derivative, or equivalently by a minimum of the pressure. As we
discuss below, the corresponding singularity of the thermodynamic potential is weaker
than at a regular spinodal line.
In Fig.~\ref{phase} the phase diagram of the NJL model in the $(T,n_q)$ plane is shown,
including the isentropic and isothermal spinodal lines. For finite $m$ the NJL model
yields a generic phase diagram with a first-order phase transition at low temperatures and
high densities and a cross-over transition at high temperatures. The critical end point
separates the cross-over from  the first-order transition. In the chiral limit, the
cross-over transition becomes a second-order phase transition. In an equilibrium
first-order phase transition, the meta-stable and unstable regions are not probed. The
states of maximum entropy, obtained by the Maxwell construction, correspond to the
coexistence of the high- and low-density states at the corresponding phase boundaries. On
the other hand, in a non-equilibrium system, where the thermal parameters are changing
sufficiently fast, the system may traverse the meta-stable and enter the unstable region,
bounded by the spinodal lines. From the thermodynamic relation
\begin{align}\label{eq6}
& & \left( \frac{\partial P}{\partial V} \right)_T = \left( \frac{\partial P}{\partial V}
\right)_S {}+ \frac{T}{C_V}\left[ \left( \frac{\partial P}{\partial T} \right)_V
\right]^2\,,
\end{align}
it is clear that the isentropic spinodal lines are  located inside  the isothermal
spinodal region and furthermore, that the two lines coincide at $T=0$. As shown in
Fig.~\ref{phase}, the isothermal spinodal lines join at the CEP, whereas the boundary of
the isentropic spinodal region remains well below the CEP. The latter property is
probably an artefact of the mean-field approximation. When fluctuations are included, the
specific heat at constant volume $C_V$ diverges at the CEP, although with a weaker
singularity than $C_P$, while in the mean-field approximation $C_V$ remains finite (see
section 3). According to (\ref{eq6}) the isentropic and isothermal volume derivatives of
the pressure are equal for $C_V\to\infty$. This implies that also $\partial P/\partial
V|_S=0$ at the CEP and that the isentropic spinodal region probably extends to the CEP.
In order to explore this question in more detail and to determine the shape of the
isentropic spinodal region, a systematic study of fluctuations is required. At a
first-order phase transition the instabilities are reflected in a convex structure of
several thermodynamic functions. Such a structure is seen in the thermodynamic pressure
and also appears in the entropy density as a function of quark density $n_q$
\begin{align}
s &=  - \frac{\partial\Omega}{\partial T}
\nonumber\\
&= - 2 N_c N_f \int\frac{d^3 p}{(2\pi)^3} \Bigl[ \ln (1 - n_f^{(+)}) + \ln (1 -
n_f^{(-)})
\nonumber\\
&\quad {}- \frac{E_f}{T}\left( n_f^{(+)} + n_f^{(-)} \right) {}+ \frac{\mu}{T}\left(
n_f^{(+)} - n_f^{(-)} \right) \Bigr]\,.
\end{align}
However, the entropy per quark, $s/n_q$, exhibits a monotonic dependence on the density.
In Fig.~\ref{sb} we show  $s/n_q$ computed at fixed temperature in the chiral limit and
for a finite quark mass.  In the chiral limit the entropy per quark shows a cusp
structure at the spinodal line on the high-density side.

\section{Susceptibilities and the critical behavior}
\label{sec:sus}

At the critical end point the fluctuations of the quark density diverge~\cite{pd,SFR:prd,
SFR:NJL,fujii,stephanov,hatta1}, while they are finite above and below the CEP if the
first order transition takes place in equilibrium. However, in non-equilibrium the system
may reach the spinodal instability. It is thus natural to explore the charge fluctuations
in the metastable and unstable regions of the first order phase transition.
\begin{figure*}
\begin{center}
\includegraphics[width=8.7cm]{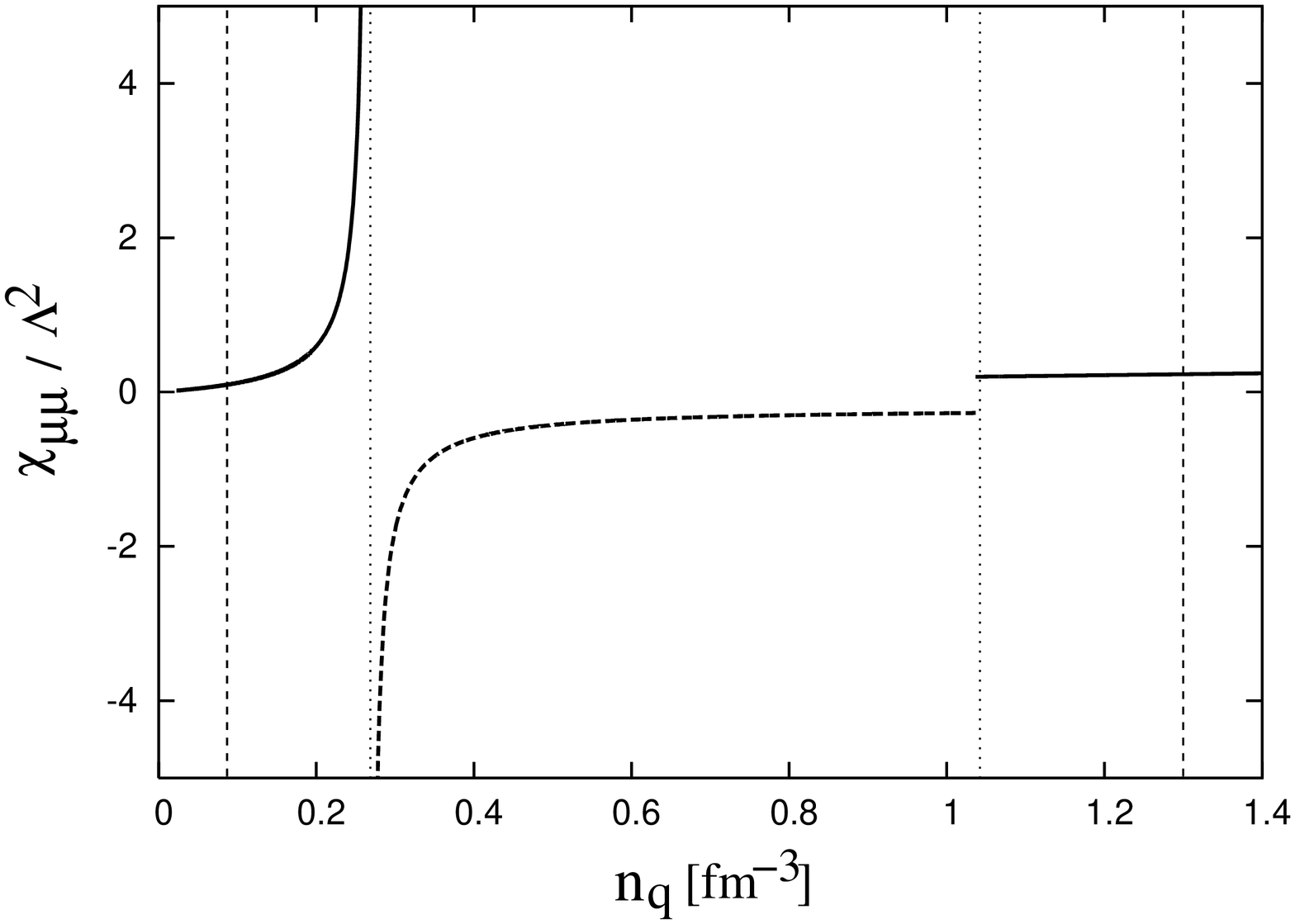}
\includegraphics[width=8.7cm]{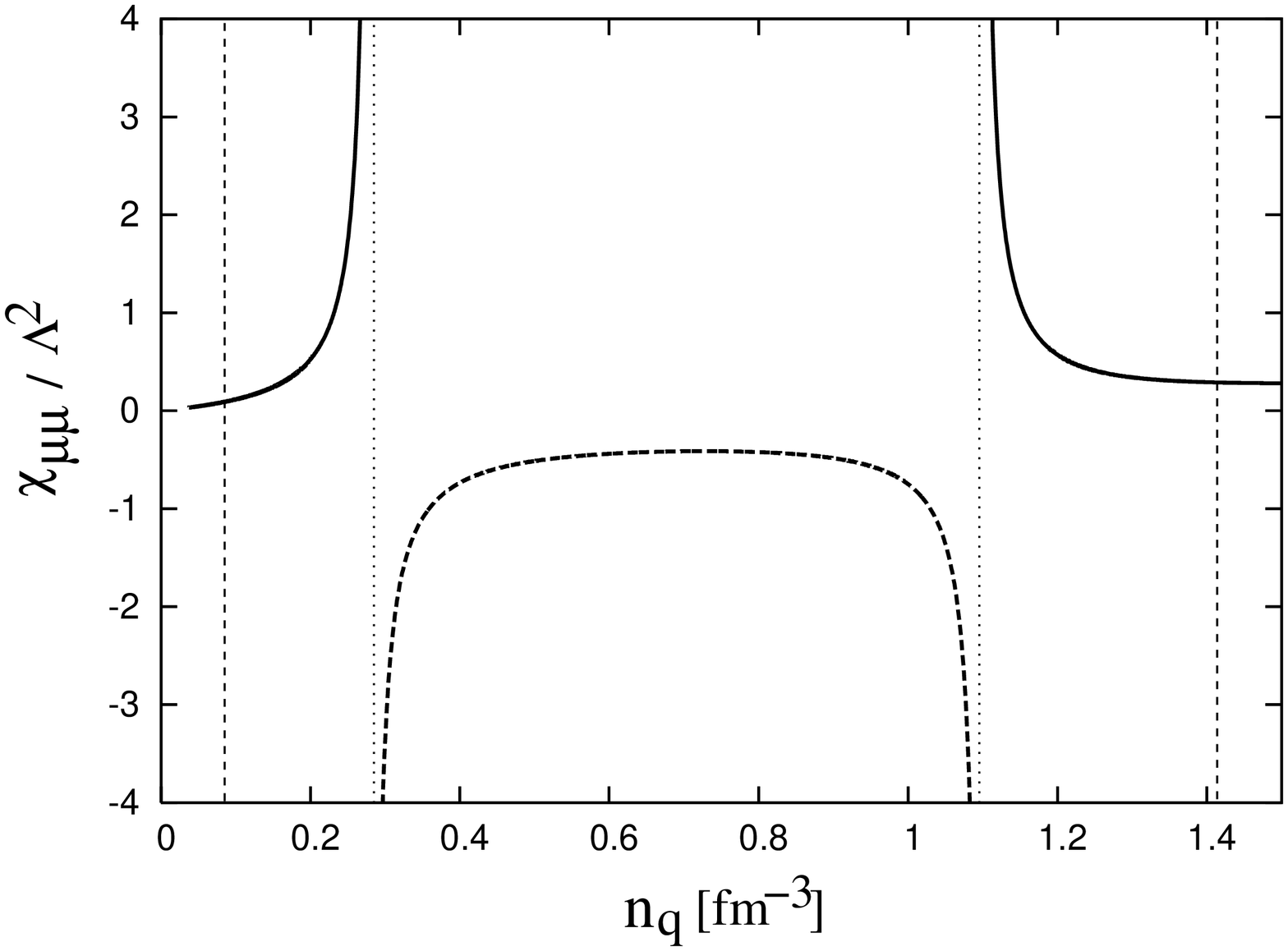}
\caption{ The $\chi_{\mu\mu}$ susceptibility at fixed temperature $T=30$ MeV
in the chiral limit (left panel) and for a finite quark mass $m=5.6$
MeV (right panel). The vertical dashed and dotted lines have the same meaning as in Fig.~\ref{mass}.}
\label{qsus}
\end{center}
\end{figure*}
In statistical systems a measure of the fluctuations of a particular quantity is provided
by the corresponding susceptibility $\chi$. This are obtained as second derivative of the
thermodynamic pressure. For the characterization of the spinodal region of the first
order chiral phase transition, the susceptibilities related to the net quark number
density $\chi_q\equiv\chi_{\mu\mu}$, electric charge density $\chi_Q$ as well as the
thermal $\chi_{TT}$ and mixed $\chi_{\mu T}$ susceptibilities are of particular interest.
These are obtained from the thermodynamic pressure (\ref{eq2}) as follows:
\begin{widetext}
\begin{align}
\chi_{\mu\mu} &= - \frac{\partial^2\Omega}{\partial\mu^2}
\nonumber\\
&= \frac{2 N_c N_f}{T}\int\frac{d^3 p}{(2\pi)^3} \Biggl[ \left( 1 -
\frac{M}{E}\frac{\partial M}{\partial\mu} \right) n^{(+)}(1 - n^{(+)})
{}+ \left( 1 + \frac{M}{E}\frac{\partial M}{\partial\mu} \right) n^{(-)}(1 - n^{(-)})
\Biggr]\,,
\\
\chi_{\mu T} &= - \frac{\partial^2\Omega}{\partial\mu\partial T}
\nonumber\\
&= \frac{2 N_c N_f}{T^2}\int\frac{d^3 p}{(2\pi)^3} \Biggl[ \left( 1 -
\frac{M}{E}\frac{\partial M}{\partial\mu} \right) (E - \mu)\,n^{(+)}(1 - n^{(+)})
{}- \left( 1 + \frac{M}{E}\frac{\partial M}{\partial\mu} \right)
(E + \mu)\,n^{(-)}(1 - n^{(-)}) \Biggr]\,,
\\
\chi_{TT} &= - \frac{\partial^2\Omega}{\partial T^2}
\nonumber\\
&= \frac{2 N_c N_f}{T^2}\int\frac{d^3 p}{(2\pi)^3} \Biggl[ \left( \frac{E - \mu}{T}
{}- \frac{M}{E}\frac{\partial M}{\partial T} \right) (E - \mu)\,n^{(+)}(1 - n^{(+)})
{}+ \left( \frac{E + \mu}{T} {}- \frac{M}{E}\frac{\partial M}{\partial T} \right)
(E + \mu)\,n^{(-)}(1 - n^{(-)}) \Biggr]\,.
\nonumber\\
\end{align}
\end{widetext}
The susceptibilities involve derivatives of the dynamical quark mass with respect
 to $\mu$ or $T$~\footnote{The above form of the mixed susceptibility is obtained
 from $\chi_{\mu T} = \partial s/\partial \mu$. An alternative equation, involving
 $\partial M/\partial T$, is obtained from $\chi_{\mu T} = \partial n_q/\partial T$.
  The equivalence of the two is obvious when the explicit expressions
   for $\partial M/\partial \mu$ and $\partial M/\partial T$ are inserted.}.
 Consequently, at the spinodal lines, where
  both $\partial M/\partial \mu$ and $\partial M/\partial T$ diverge (see Fig.~\ref{mass}),
one expects singularities in these susceptibilities. In Fig.~\ref{qsus} we show the quark
number susceptibility as a function of $n_q$ at a fixed temperature $T=30$ MeV. As
expected, it is singular at the spinodal points. However, in the chiral limit the
susceptibility exhibits only a weak singularity at the high-density point, corresponding
to the cusp in the pressure in Fig.~\ref{pv}. At this point the susceptibility is
discontinuous, changing from a finite negative to a finite positive value, while at the
low density spinodal point the susceptibility diverges, corresponding to very large
fluctuations of the net quark density. Moreover, for a finite current quark mass, the
susceptibility diverges at both spinodal points. In both cases the quark number
susceptibility is negative in the spinodal region,
 indicating a mechanical instability of the system.

The critical properties of the net quark number susceptibility seen at the isothermal
 spinodal lines are not there along the isentropic spinodal trajectories.
There,  the susceptibilities are finite and continues when the system passes through the
metastable region. The above behavior of $\chi_{\mu\mu}$  appears as a direct consequence
of the thermodynamic relations
\begin{align}\label{eq7}
& \left( \frac{\partial P}{\partial V} \right)_T = -
\frac{n_q^2}{V}\frac{1}{\chi_{\mu\mu}}\,,
\\
&  \left( \frac{\partial P}{\partial V} \right)_S = - \frac{n_q^2}{V}\frac{\chi_{TT} -
\frac{2 s}{n_q}\chi_{\mu T} {}+ \left( \frac{s}{n_q} \right)^2 \chi_{\mu\mu}}
{\chi_{\mu\mu}\chi_{TT} - \chi_{\mu T}^2}\,,\label{eq8}
\end{align}
that connect the pressure derivatives with the susceptibilities
$\chi_{xy}=-\partial^2\Omega /\partial x\partial y$.

Along the  isothermal spinodal lines the pressure derivative  in Eq.~(\ref{eq7})
vanishes. Thus, for non vanishing density $n_q$ its fluctuations have to diverge. In
addition, since the pressure derivative ${\partial P}/{\partial V}|_T$ changes its sign
when crossing the spinodal trajectories, there is a corresponding sign change of the
divergence in $\chi_{\mu\mu}$ as seen in Fig.~\ref{qsus}.
A similar behavior as   for $\chi_{\mu\mu}$ is also expected  for electric charge
$\chi_Q$, strangeness $\chi_S$ or isovector $\chi_I$ fluctuations if their charge
densities are finite.

In heavy ion collisions the isospin asymmetry is usually negligible at high energy and
the strangeness density vanishes due to strangeness neutrality in the initial state.
Thus, here not all chemical potentials are thermodynamically independent. The net quark
 number $\chi_q = - \partial^2\Omega / \partial\mu_q^2$ and
  the isovector $\chi_I = - \partial^2\Omega / \partial\mu_I^2$ susceptibilities are
   related with the electric charge fluctuations $\chi_Q$ as
\begin{equation}
\chi_Q=\frac{1}{36}\chi_q + \frac{1}{4}\chi_I +
\frac{1}{6} \frac{\partial^2P}{\partial \mu_q\partial\mu_I}\,,
\label{elec}
\end{equation}
For isospin symmetric system  the  derivative of the pressure vanishes and all relevant
susceptibilities are linearly dependent. Clearly, since $\chi_q$ diverges on the isothermal spinodals
 and since $n_I\simeq 0$ is a good approximation in heavy ion collisions, thus the electric charge
 fluctuations $\chi_Q$ should show similar critical behavior to that seen in $\chi_q$.
On  the isentropic spinodals the quark number fluctuations are finite. This is also valid for all
 mixed susceptibilities appearing in  Eq.~(\ref{eq8}). The isentropic condition (\ref{eq5}) holds
since  the numerator in (\ref{eq8}) vanishes. The exception is the point at  $T=0$ where
the isothermal and isentropic spinodal conditions are equivalent since   the entropy
density and the  $\chi_{\mu T}$ vanish  at $T=0$. The thermal $\chi_{TT}$ and mixed
$\chi_{\mu T}$ susceptibilities along the isothermal spinodals show similar critical
behavior as found for $\chi_{\mu\mu}$ in Fig.~\ref{qsus}. When going beyond the mean
field dynamics, the isothermal and the isentropic spinodal curves could also coincide at
the CEP. Thus, from Eqs. (\ref{eq7}) and (\ref{eq8}) it is clear that all
susceptibilities should diverge at the CEP in this case.

For   equilibrium first order
phase transition the properties of  charge fluctuations are different from those seen in
Fig.~\ref{qsus}. There, under the mean field approximation, the charge susceptibilities
are positive and exhibit a discontinuity at the transition point \cite{hatta,our}. This
discontinuous structure is eventually converted to a cusp if the quantum correction are
included~\cite{hatta,bj}. However, the essential difference is that the susceptibilities
in the equilibrium transition are finite whereas they are divergent if spinodal phase
separation appears.

\begin{figure*}
\begin{center}
\includegraphics[width=8.7cm]{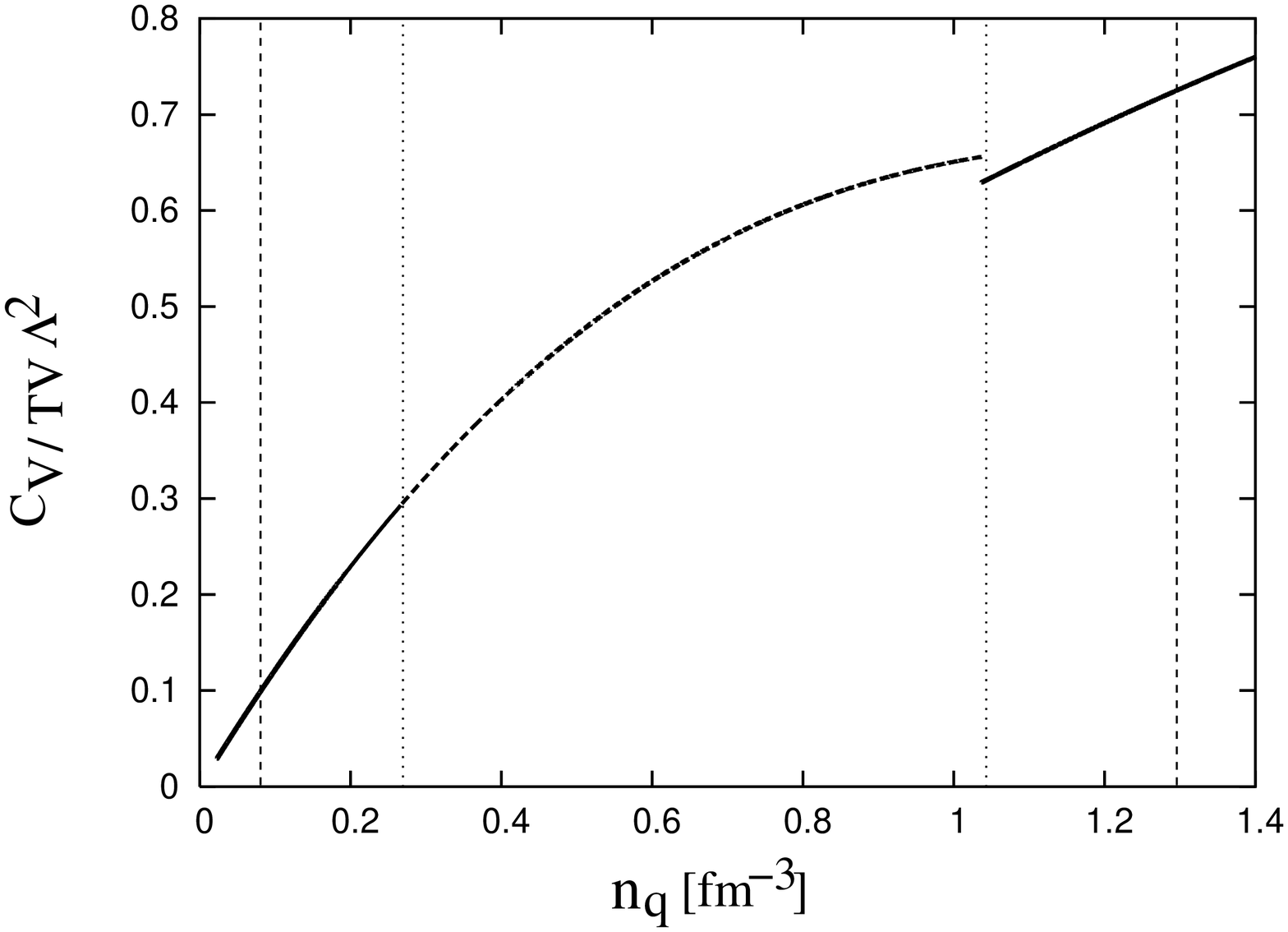}
\includegraphics[width=8.7cm]{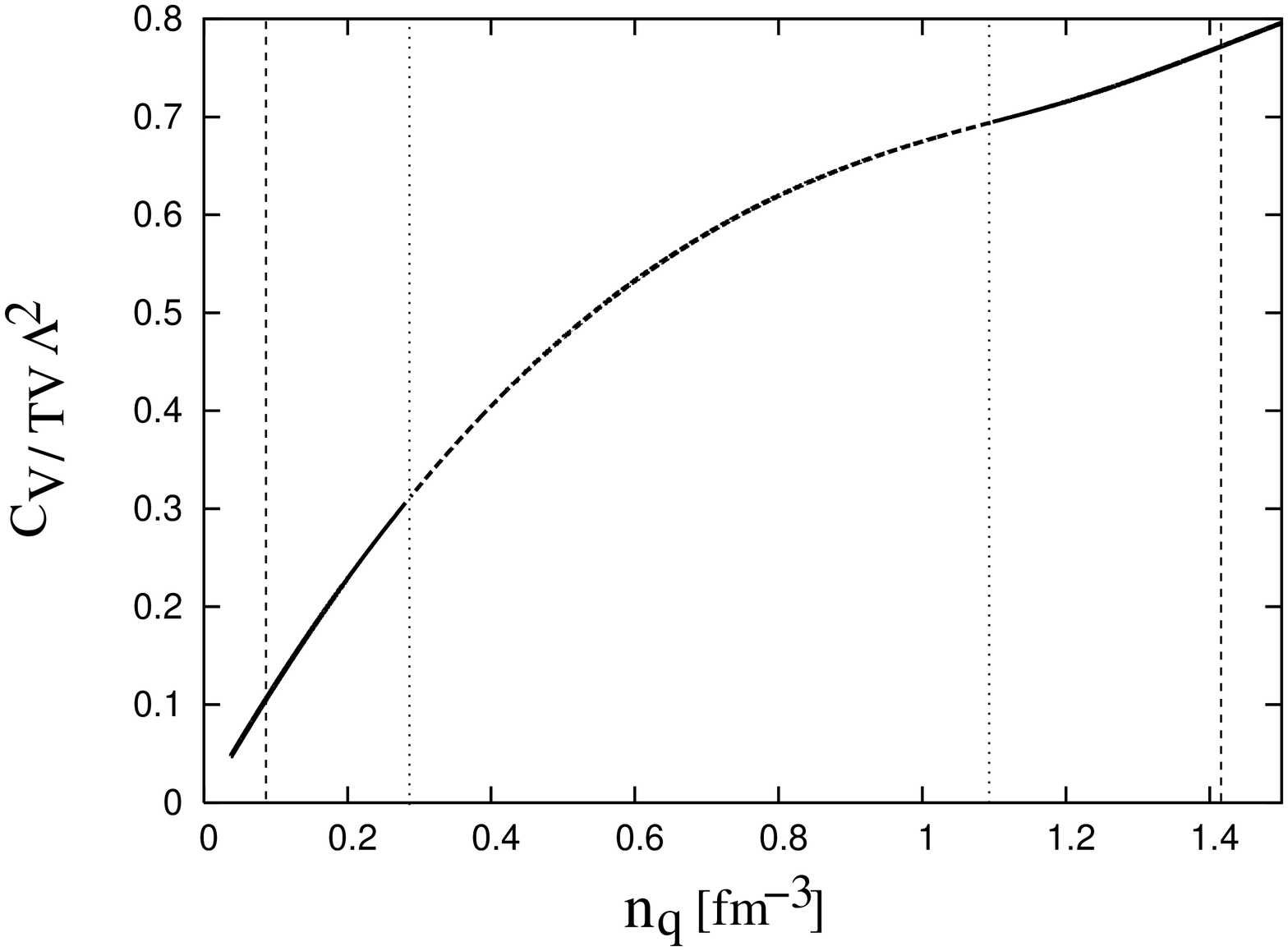}
\caption{ The specific heat $C_V$ at constant volume and  at fixed temperature $T=30$ MeV
in the chiral limit (left panel) and for a finite quark mass $m=5.6$ MeV (right panel).
The vertical dashed-lines indicate the location of the first order phase transition.
The dotted-lines show the position of the isothermal spinodal points.}
\label{cv}
\end{center}
\end{figure*}
\begin{figure*}
\begin{center}
\includegraphics[width=8.7cm]{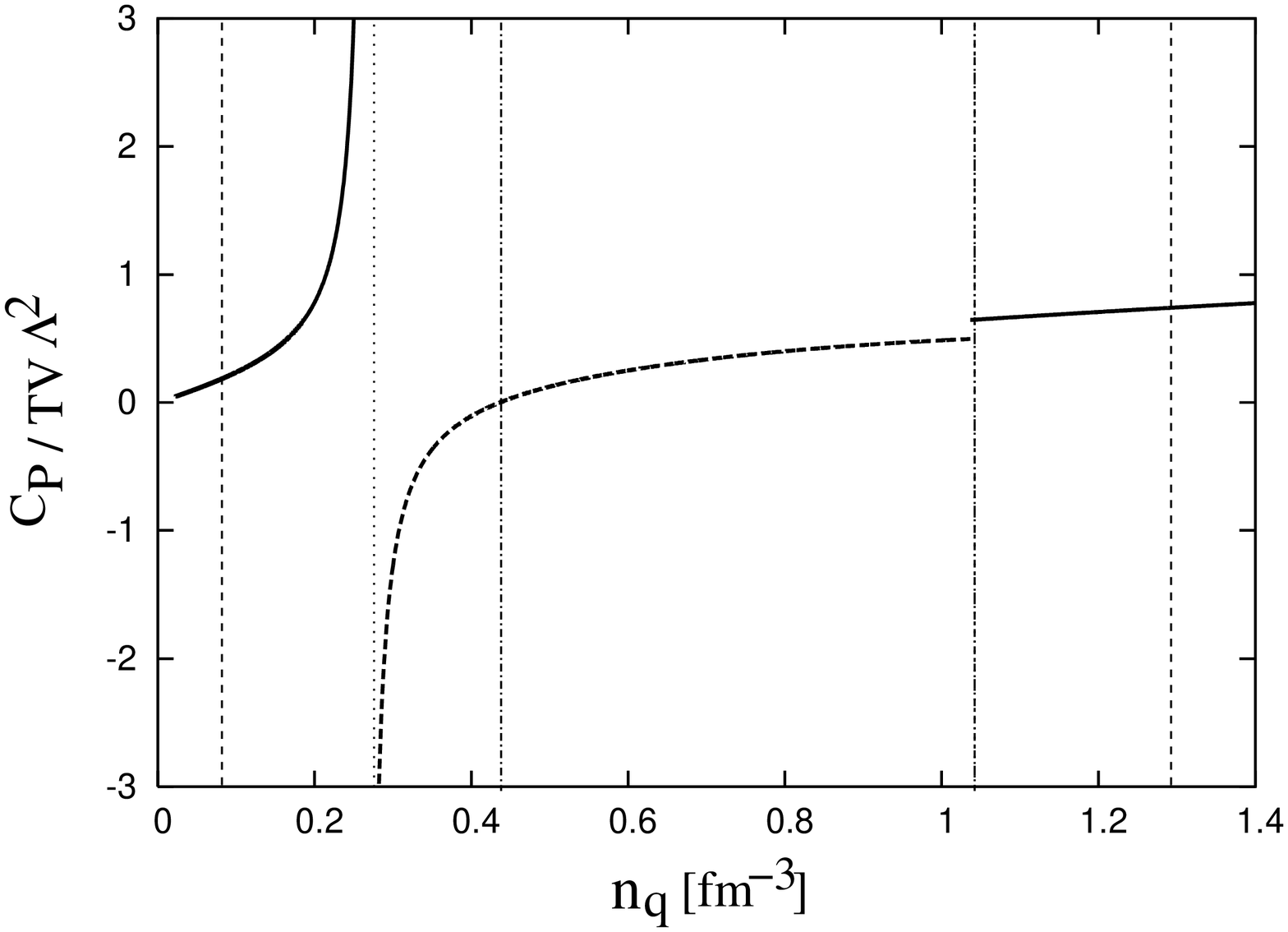}
\includegraphics[width=8.7cm]{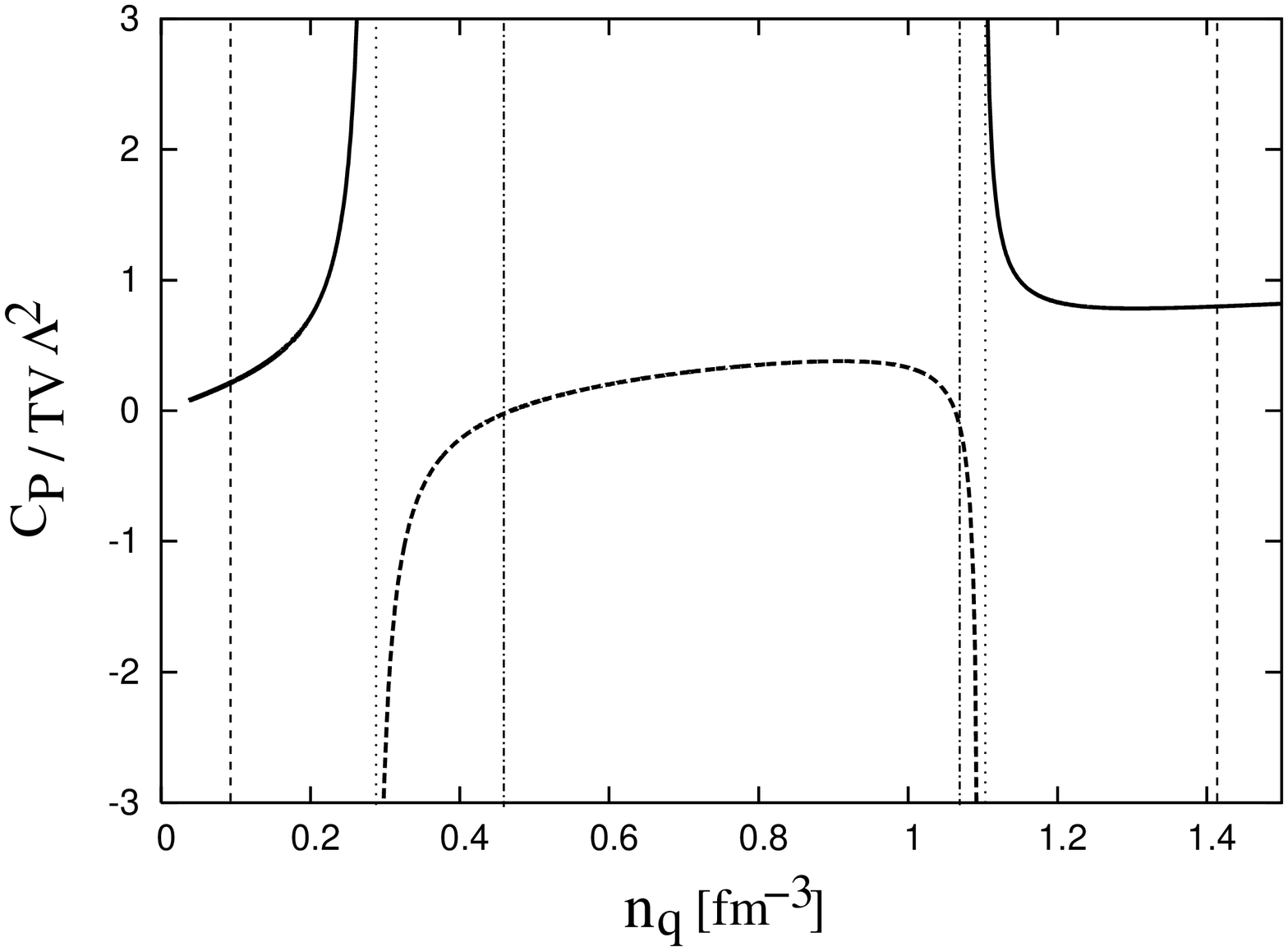}
\caption{ The specific heat $C_P$ at constant pressure and  at fixed temperature $T=30$
MeV in the chiral limit (left panel) and for a finite quark mass $m=5.6$ MeV (right panel).
The vertical dashed-lines indicate the location of the first order phase transition.
The dotted/dashed-dotted lines show the position of the isothermal/isentropic
spinodal points.}
\label{cp}
\end{center}
\end{figure*}

The rate of change in entropy with respect to temperature at constant volume or pressure
gives the specific heat expressed as
\begin{align}
C_V
&= T \left( \frac{\partial S}{\partial T} \right)_V
= TV \left[ \chi_{TT} - \frac{\chi_{\mu T}^2}{\chi_{\mu\mu}} \right]\,,
\label{eq:cv}
\\
C_P
&= T \left( \frac{\partial S}{\partial T} \right)_P
= TV \left[ \chi_{TT} - \frac{2 s}{n_q}\chi_{\mu T}
{}+ \left( \frac{s}{n_q} \right)^2 \chi_{\mu\mu} \right]\,.
\label{eq:cp}
\end{align}
From Eqs.~(\ref{eq8}), (\ref{eq:cv}) and (\ref{eq:cp}) 
it is clear that the ratio of specific heats $C_P/C_V$
satisfies a well known relation
\begin{align}
\left( \frac{\partial P}{\partial V} \right)_S= {{C_P}\over {C_V}} \left( \frac{\partial
P}{\partial V} \right)_T.
\end{align}
In Fig.~\ref{cv} we show the specific heat $C_V$ at fixed temperature. One sees in the
chiral limit a discontinuity at the right-handed spinodal branch which comes from a
similar jump in $\chi_{xy}$. At the left-handed spinodal branch $C_V$ is continuous and
finite. Each susceptibility $\chi_{xy}$ diverges at the spinodal point, however such
singularities are canceled out in Eq.~(\ref{eq:cv}). The above is also valid for finite
quark masses, thus  $C_V$ continuously changes at both spinodals.

On the other hand, as  seen in Fig.~\ref{cp},  the $C_P$  diverges  at the isothermal
spinodals  because there is  no  cancellation of the singularities among $\chi_{xy}$ in
Eq.~(\ref{eq:cp}). It has been argued that the negative specific heat could be a signal
of the liquid-gas phase transition~\cite{chomaz} and its occurrence has recently been
reported as the first experimental evidence for such an anomalous behavior in low-energy
nuclear collisions \cite{experiment,exp1}.

From Eqs.~(\ref{eq8}) and (\ref{eq:cp}) one finds that $C_P$ vanishes along the
isentropic spinodal lines. However,  this is not the case in the chiral limit. Here the
pressure shows a cusp, thus its  derivative is not defined. Consequently,  $\partial
P/\partial V|_S$ is non-vanishing at this  isentropic spinodal line  and  $C_P$ remains
finite.

\begin{figure}
\begin{center}
\includegraphics[width=8.7cm]{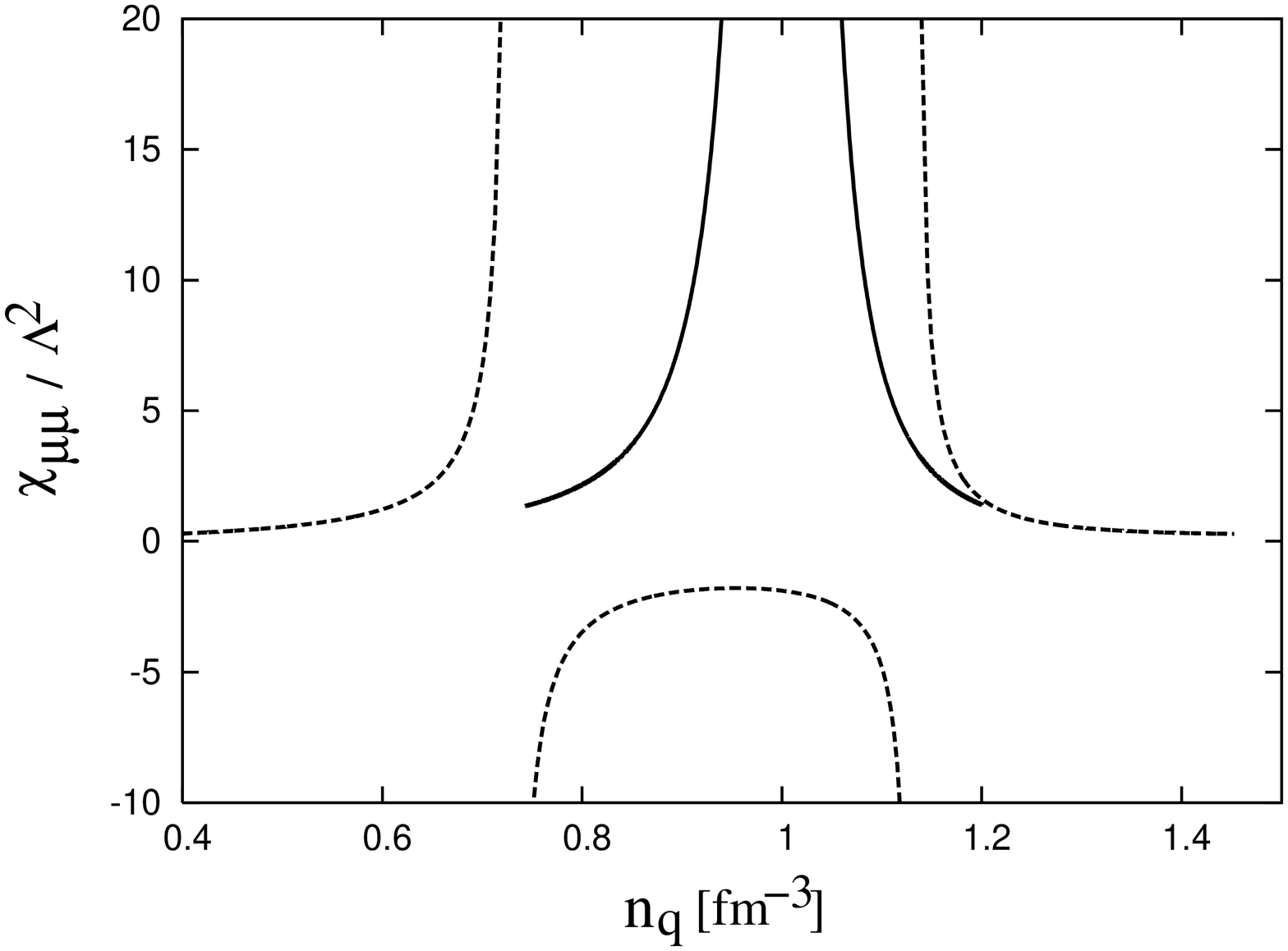}
\caption{ The  net quark number density fluctuations $\chi_{\mu\mu}/\Lambda^2$ normalized
to momentum cut-off as a function of the quark  number density $n_q$. The full-line
represents the fluctuations at the CEP at constant temperature $T_{\rm CEP}\simeq 80$ MeV.
The dashed-line shows the  $\chi_{\mu\mu}$ at fixed $T=70$ MeV corresponding to the 1st
order transition.
 }
\label{evolv}
\end{center}
\end{figure}
\begin{figure*}
\begin{center}
\includegraphics[width=8.7cm]{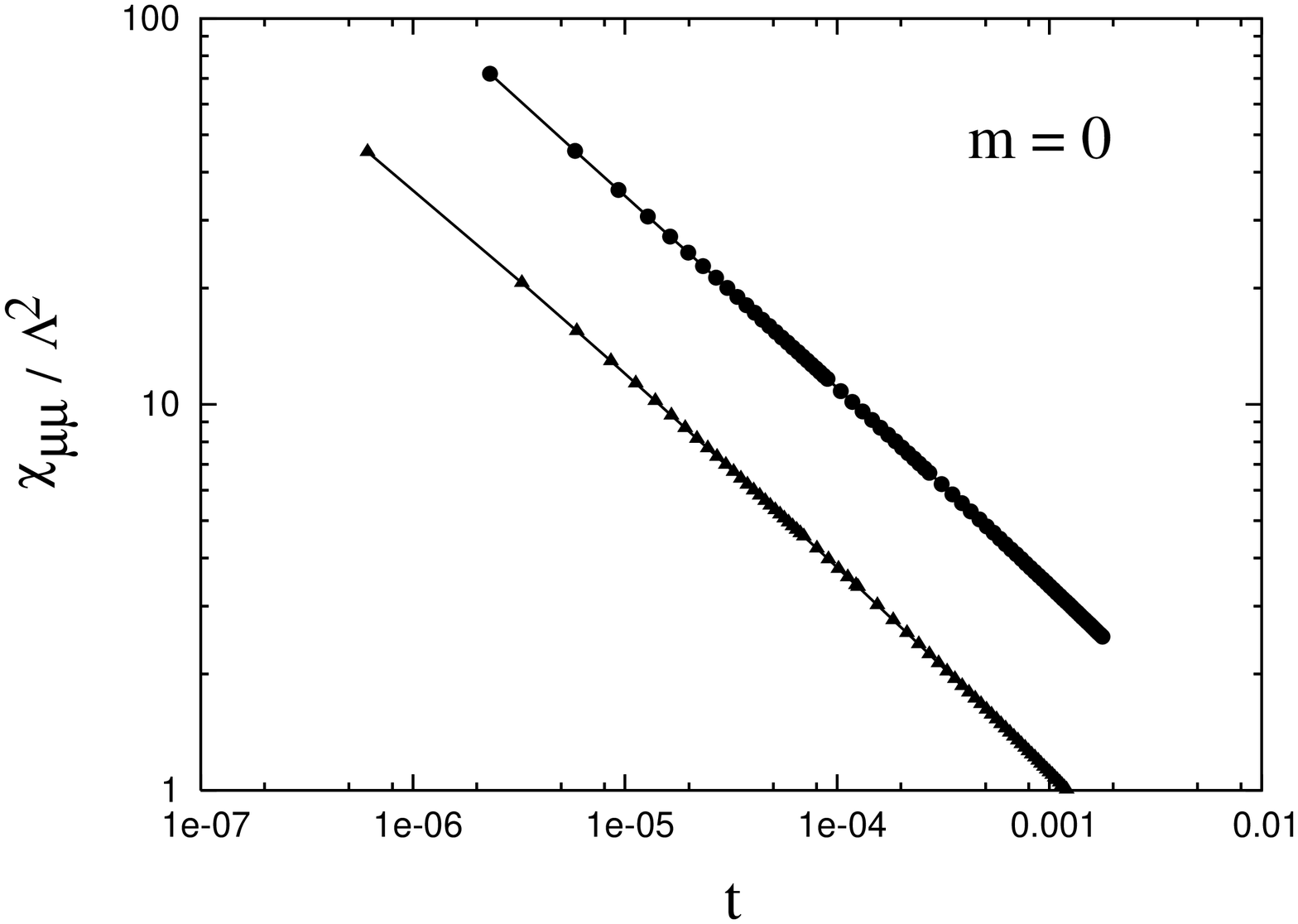}
\includegraphics[width=8.7cm]{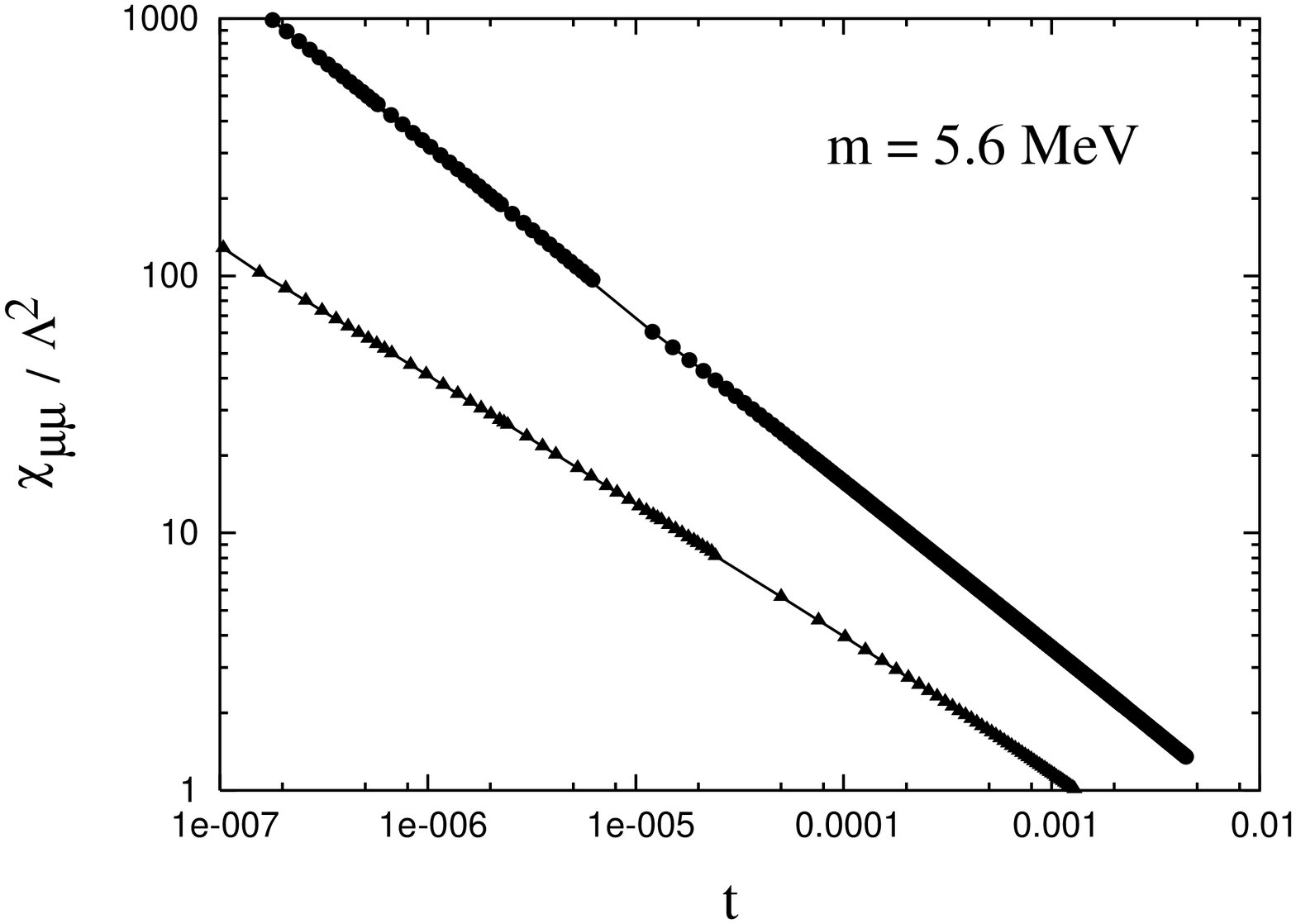}
\caption{The net quark number susceptibility in the vicinity of the TCP (left panel) and
CEP (right panel) as a function of the reduced quark chemical  potential
$t=(\mu-\mu_c)/\mu_c$ at fixed $T$. The filled circle ($\bullet$) denotes the results
calculated at $T=T_{\rm TCP/CEP}$ and the filled triangle ($\blacktriangle$) at $T=30$
MeV $< T_{\rm TCP/CEP}$ corresponding to the first order transition. } \label{exponent}
\end{center}
\end{figure*}
\begin{figure*}
\begin{center}
\includegraphics[width=8.7cm]{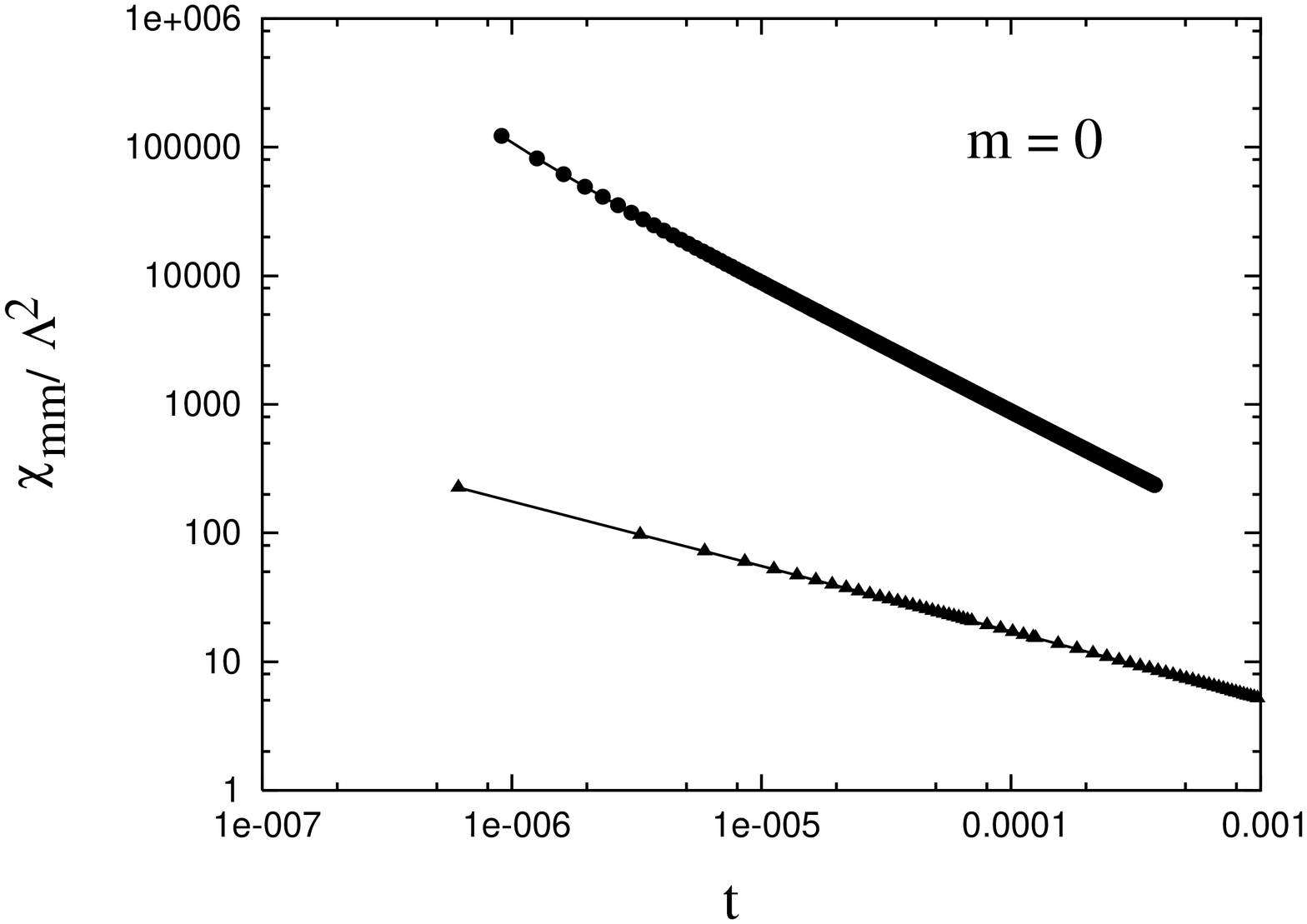}
\includegraphics[width=8.7cm]{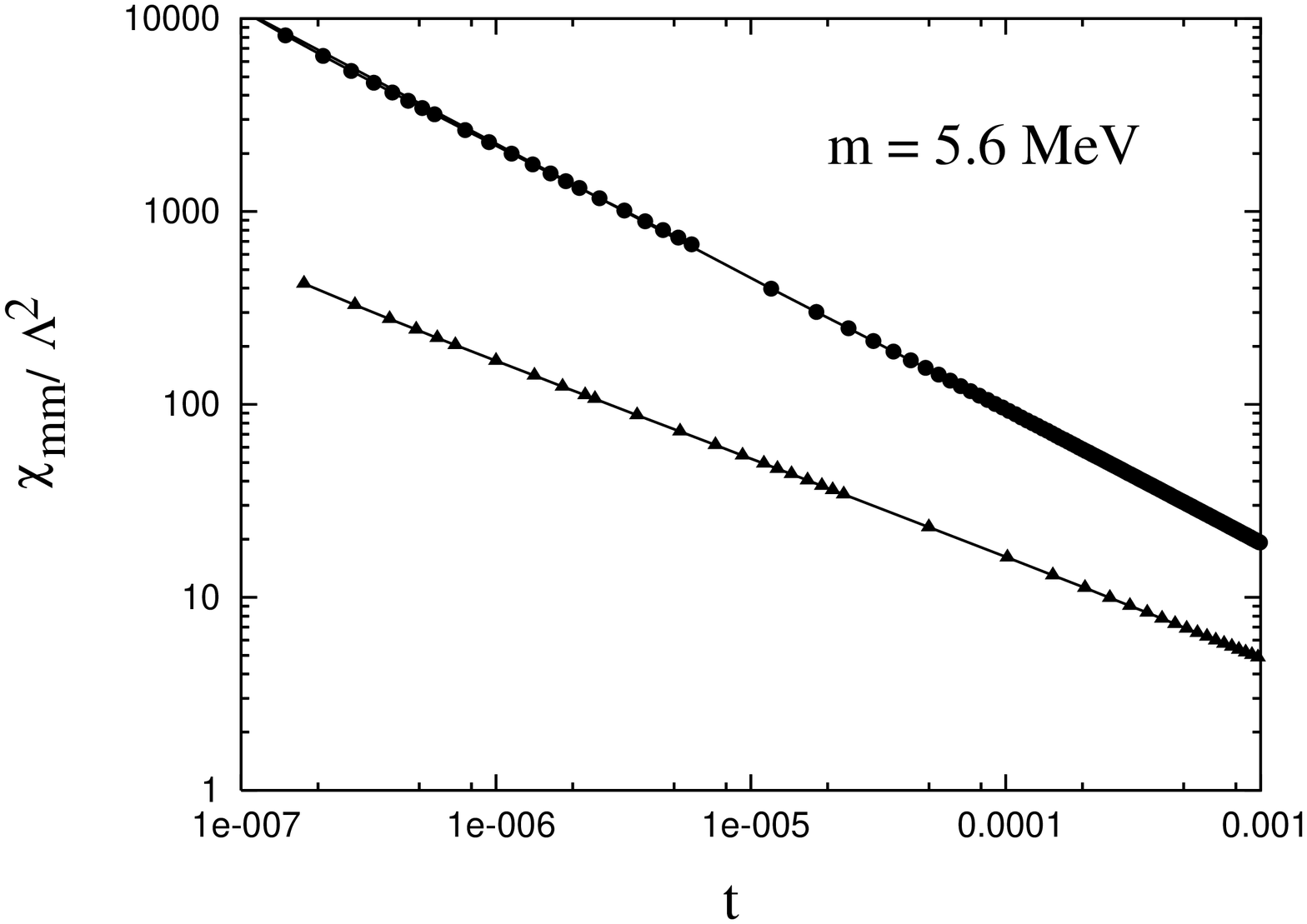}
\caption{The chiral  susceptibility in the vicinity of the TCP and spinodals in the
chiral limit  (left panel) and at finite quark mass  (right panel) as a function of the
reduced quark chemical potential $t=(\mu-\mu_c)/\mu_c$ at fixed $T$. The filled circle
($\bullet$) denotes the results calculated at $T=T_{\rm TCP/CEP}$ and the filled triangle
($\blacktriangle$) at $T=30$ MeV $< T_{\rm TCP/CEP}$ corresponding to the first order
transition. } \label{cs1}
\end{center}
\end{figure*}

At the CEP the fluctuations of  charge densities that couple to the massless modes of the
sigma field are divergent. Fig.~\ref{evolv} shows the  $\chi_{\mu\mu}$ near the CEP and
at fixed temperature corresponding to the first order transition. From this figure it is
clear that when approaching the CEP from the side of the first order transition the
region of instabilities shrinks and at the CEP disappears completely. Consequently, the
CEP singularity in $\chi_{\mu\mu}$ appears from the matching of the two positive singular
branches of $\chi_{\mu\mu}$. The strength of the singularity is controlled by the
critical exponent $\gamma$ shown in Fig.~\ref{exponent}. In the chiral limit the exponent
at the isothermal spinodal line is identical to that at the TCP. From numerical analysis
we find that  $\chi_{\mu\mu}\sim (\mu-\mu_c)^{-\gamma}$, with $\gamma=1/2$. However, for
finite quark masses we find $\gamma=2/3$ at CEP, in  agreement with that expected in the
mean-field approximation~\cite{hatta,bj}.
From the numerical studies of $\chi_{\mu\mu}$ at spinodals we found that $\gamma=1/2$
independently of  the value of the quark mass. Thus, the singular properties at the
TCP/CEP are the remnant of divergent behavior of the $\chi_{\mu\mu}$ at the first order
chiral phase transition~\cite{SFR:spinodal}. Clearly, the critical exponents are
renormalized by quantum fluctuations \cite{bj}.  However,  the smooth evolution of the
singularity from the spinodal lines to the CEP is expected to be generic.

In the next section we disuse how the critical exponents of quark susceptibilities
along the spinodal lines can be analytically justified within the Ginzburg-Landau theory.

\section{Ginzburg-Landau effective theory and critical exponents}
\label{sec:GL}
In the mean filed approximation and near the chiral phase transition the thermodynamic
potential obtained in  QCD-like chiral models can be described by the Ginzburg-Landau
effective theory ~\cite{LL}. Consequently, the mean field critical exponents at the
chiral phase boundary can be derived analytically. In  earlier studies within this
approach the critical exponents of different susceptibilities were obtained assuming that
the system appears in  equilibrium~\cite{hatta,SFR:NJL}. In the following we extend the
Ginzburg-Landau formalism to account for  spinodal phase decomposition in the meta-stable
regime of the 1st order transition.

According  to  the Ginzburg-Landau theory, close  to the phase boundary, the
thermodynamic potential may be expanded in power series of the order parameter $M$~\cite{LL}:
\begin{equation}\label{GL}
\Omega (T,\mu;M)=\Omega_0-h M+\frac{1}{2} a_2 M^2+\frac{1}{4} a_4 M^4
{}+ \frac{1}{6} a_6 M^6\,,
\end{equation}
where $h M$ is a term that explicitly breaks the symmetry $M \leftrightarrow -M$.
 In QCD this corresponds to the quark mass term.
The gap equation
\begin{equation}\label{gap}
\Omega^\prime (M)= - h + a_2 M + a_4 M^3 + a_6 M^5 = 0
\end{equation}
yields the values of the order parameter that extremize the thermodynamic potential. Consider first
the case without explicit symmetry breaking, i.e. $h=0$. Then, at a second-order phase
transition $a_2=0$ and $a_4>0$, while at a first order transition $a_2>0$, $a_4<0$.
 The parameter $a_6$ should be positive for stability. Clearly, the TCP can be identified by $a_2=a_4=0$.
If $h\neq 0$ and $a_4>0$, a change of sign of $a_2$ corresponds to a cross over transition.
We note that in this case it is not possible to uniquely define the position of the transition,
 since the order parameter $M$ is always non-zero. The endpoint of the first order transition
 is then called a critical end point (CEP).
In order to find the spinodal lines, we need also the second derivative of $\Omega$
\begin{equation}\label{second-D}
\Omega^{\prime\prime}(M)=a_2+3a_4 M^2+5a_6 M^4 \, .
\end{equation}
The zeroes of the second derivative, $\Omega^{\prime\prime}(M_0)=0$, are given by
\begin{equation}\label{sol2}
M_0^2=-\frac{3 a_4}{10 a_6} \pm \frac{1}{10 a_6}\sqrt{9 a_4^2-20 a_2 a_6}\, .
\end{equation}
Depending on the values of the parameters, the solutions correspond to inflexion points, or, if $M_0$
 also solves the gap equation,
 to a point on a spinodal line.
 We expand the thermodynamic potential around a zero of the second derivative
\begin{equation}\label{exp-m}
\Omega(m)=\tilde{\Omega}_0+A_1 m + \frac{A_2}{2} m^2 +\frac{A_3}{3} m^3 + \frac{A_4}{4} m^4\, ,
\end{equation}
where $m=M-M_0$. One finds
\begin{eqnarray}\label{coeffs}
A_1&=&-h+a_2 M_0 +a_4 M_0^3 + a_6 M_0^5\nonumber\\
A_2&=&a_2 + 3 a_4 M_0^2 + 5 a_6 M_0^4=0\nonumber\\
A_3&=&3 a_4 M_0 + 10 a_6 M_0^3 = \pm M_0 \sqrt{9 a_4^2-20 a_2 a_6}\\
A_4&=&a_4 + 10 a_6 M_0^2 = -2a_4 \pm\sqrt{9 a_4^2-20 a_2 a_6}\, .\nonumber
\end{eqnarray}

At a spinodal line the gap equation is satisfied, i.e., $A_1=0$. Thus, close to a spinodal line,
\begin{eqnarray}\label{close-sp}
A_1&=&a_1(T-T_0) + b_1(\mu - \mu_0)\nonumber\\
A_2&=&a_2(T-T_0) + b_2(\mu - \mu_0)\, ,
\end{eqnarray}
while the coefficient $A_3$ is non-zero, and can be taken to be constant over the small range of
temperatures and chemical potentials considered.
 In the gap equation for a point close to the spinodal line $\mu\simeq \mu_0$ (for definiteness we set
 $T=T_0$),
\begin{equation}\label{gap2}
A_1 + A_2 m + A_3 m^2 + A_4 m^3 = 0\, ,
\end{equation}
we assume that $m\sim \pm|\mu-\mu_0|^\alpha$,
with~\footnote{The assumption  $|\alpha|>1$ does not yield a
consistent solution of (\ref{gap2}).} $0<\alpha <1$.
For small $|\mu-\mu_0|$, the leading terms of (\ref{gap2}) are then $A_1+A_3 m^2 = 0$,
 yielding $\alpha=1/2$. At the critical end point also $A_3=0$,
 so the leading terms in the gap equation are $A_1+A_4 m^3 = 0$, and correspondingly $\alpha=1/3$.
The singular part of the baryon number susceptibility is given by
\begin{equation}\label{sing}
\chi_{\mu\mu}^{sing}=-\frac{\partial^2\Omega}{\partial\mu^2}=-b_1\frac{\partial m}
{\partial \mu}\sim |\mu-\mu_0|^{-\gamma}\, ,
\end{equation}
with $\gamma = 1/2$ at the spinodal lines and $\gamma=2/3$ at the CEP. Thus, the
singularities of the two spinodal lines conspire to yield a somewhat different critical
exponent at the CEP.
In the chiral limit, $\alpha=1/2$ at the spinodal lines and the corresponding critical
 exponent is $\gamma=1/2$. Finally, at the TCP $\alpha=1/4$,
  but the critical exponent is unchanged, since here $M=0$ and
\begin{equation}\label{TCP-exp}
\chi_{\mu\mu}\sim M\frac{\partial M}{\partial \mu}\sim\frac{\partial M^2}{\partial \mu}
\sim |\mu-\mu_0|^{-1/2}\, .
\end{equation}
Thus, in the chiral limit the critical exponents of the baryon number susceptibility at
the spinodal lines and at the TCP are identical, while for finite quark masses the
corresponding exponents differ.
The critical  and the universal properties  at the phase boundary can be also studied
considering the fluctuations of the order parameter.  They  are characterized by the
chiral susceptibility, $\chi_{mm}$ defined as the second order derivative of
thermodynamic potential with respect to the quark mass
\begin{equation}
\chi_{mm}=-{{\partial^2\Omega}\over{\partial m^2}}.
\end{equation}
Figure \ref{cs1}  shows the change of the chiral susceptibility near the critical end
point  and along the spinodal lines in the chiral limit and for the finite value of the
quark masses. In the chiral limit the behavior of $\chi_{mm}$ as a function of
$t=|(\mu-\mu_c)/\mu_c|$ differs from that shown in Fig. \ref{exponent} for the net quark
number density fluctuations $\chi_{\mu\mu}$. First, the critical exponents at TCP and
along spinodals do not coincide. Second, the critical exponent $\gamma$  at the TCP is
now equal to unity instead of $\gamma=1/2$ whereas at spinodals it still remains the same
as for $\chi_{\mu\mu}$. On the other hand, at finite quark mass the critical exponents at
the CEP and at the spinodals for  $\chi_{mm}$ are the same as found for $\chi_{\mu\mu}$.
The above properties are to be expected in mean field dynamics because of the following
relation
\begin{equation}
\chi_{\mu\mu}\simeq \chi_{\mu\mu}^{reg} +M^2\chi_{mm}, \label{cseq}
\end{equation}
that connects  $\chi_{\mu\mu}$ with  $\chi_{mm}$ and the dynamical quark mass $M$.  The
first term $\chi^{reg}$ corresponds to  a non-singular contribution to the quark number
fluctuations.
In the chiral limit, when approaching the TCP along any paths being not asymptotically
tangential to the critical line the order parameter $M^2\sim t^{1/2}$ and at the O(4)
line $M^2\sim t^{1}$ \cite{SFR:prd}. The corresponding exponents of $\chi_{mm}$ are
$\gamma=1$, what explains through  Eq. (\ref{cseq}) the critical exponents found in the
Fig.~\ref{cs1}-left. For the first-order transition in the chiral limit and for the finite 
quark mass the $M^2$ is always finite, consequently the critical exponents of $\chi_{\mu\mu}$ 
and $\chi_{mm}$ are identical as seen in Fig.~\ref{cs1}.


\section{Summary and conclusions}
\label{sec:sum}
Considering deviations from an idealized equilibrium picture of the first order chiral
phase transition by including spinodal instabilities, we have explored the critical
behavior of the fluctuations of conserved charges and the specific heat in the NJL model
in mean field approximation.  We have studied the properties these observables in the
chiral limit and for finite vale of the quark chemical potential. We showed that the
charge fluctuations diverge along the isothermal spinodal lines of the first order chiral
phase transition independently on the value of quark masses. Accompanied by the
singularities of the susceptibilities, the specific heat at constant pressure also
diverges. This is in contrast to an equilibrium transition where the susceptibilities are
finite and only at the TCP/CEP diverge. This  implies that large fluctuations of the
density would be a signal not only for the existence of the CEP or TCP  but also of a
non-equilibrium first order transition.  The above  is especially relevant to the current
and future heavy-ion experiments when considering a possible   signal for  a first order
chiral phase transition.

We also showed that in the chiral limit the critical exponents of the net quark number
susceptibility $\chi_{\mu\mu}$ are identical at the TCP and at the isothermal spinodals.
In the mean field approximation the critical properties of the net quark number
fluctuations were found to be   governed by a common critical exponent $\gamma=1/2$. On
the other hand at finite quark masses the exponents are different indicating a change of
the universality class. At the spinodals the critical exponents for $m\neq 0$ stays to be
$\gamma=1/2$. However, at the CEP it  changes from  $\gamma=1/2$ at $m= 0$ to
$\gamma=2/3$ for $m\neq 0$. These values of the critical exponents were calculated
numerically in the NJL model and also confirmed by the analytic studies of an effective
Ginzburg-Landau theory. The above results indicates that the singularities of
$\chi_{\mu\mu}$ at the TCP and CEP  are a remnant of divergent fluctuations along the two
spinodal branches.

The calculations were performed within the NJL model under mean field dynamics. However,
the appearance of the singularities in the quark fluctuations and specific heat comes
from the thermodynamic relations which connect the pressure derivatives with different
charge susceptibilities. Thus, divergent properties of charge susceptibilities and the
specific heat in the presence of spinodal instabilities are quite general,  being
independent on the specific  formulation of the chiral Lagrangian. This is also the case
for the fluctuation critical exponents at the spinodals and their relation to the
critical properties at TCP and CEP that should be valid for all QCD like chiral models
under mean field approximation. It is a further challenge to verify the influence of the
quantum and thermal fluctuations  on the critical  medium properties  in the presence of
spinodal phase separation.

\section*{ Acknowledgments}
We acknowledge stimulating  discussions with J. Randrup, M. Stephanov and S. Toneev. The
work of B.F. and C.S. was supported in part by the Virtual Institute of the Helmholtz
Association under the grant No. VH-VI-041. 
C.S. also acknowledges partial support by DFG cluster of excellence
``Origin and Structure of the Universe''.
K.R. acknowledges partial support of the Polish Ministry of National
Education (MENiSW) and DFG under the "Mercator program".


\end{document}